\documentclass{article}

\usepackage[preprint]{neurips_2026}

\usepackage[utf8]{inputenc} 
\usepackage[T1]{fontenc}    
\usepackage{hyperref}       
\usepackage{url}            
\usepackage{booktabs}       
\usepackage{amsfonts}       
\usepackage{nicefrac}       
\usepackage{microtype}      
\usepackage{xcolor}         

\usepackage{caption}
\usepackage{amssymb}
\usepackage{mathtools}
\usepackage{amsthm}
\usepackage{amsmath,amsfonts,bm}

\usepackage{hyperref}
\usepackage{url}
\usepackage{graphicx}
\usepackage{subfigure}
\usepackage{booktabs} 
\usepackage{diagbox}
\usepackage{multirow}
\usepackage[mathscr]{eucal}

\usepackage{colortbl}

\newcommand\ccg[1]{\cellcolor{yellow!30!red!30}{#1}} 
\newcommand\ccw[1]{\cellcolor{red!15}{#1}}         
\DeclareMathOperator{\diag}{diag}

\usepackage[capitalize,noabbrev]{cleveref}

\theoremstyle{plain}
\newtheorem{theorem}{Theorem}[section]

\newtheorem{lemma}[theorem]{Lemma}

\theoremstyle{definition}

\theoremstyle{remark}

\title{Towards a Theoretical Understanding of Two Tower Recommendation Models}

\author{%
  Amit Kumar Jaiswal \\
  Indian Institute of Technology (BHU)\\
  Varanasi, India 221005\\
  \texttt{amit.chr@iitbhu.ac.in,amitkumarj441@gmail.com} \\
}

\begin{document}

\maketitle

\begin{abstract}
Production-grade recommender systems rely heavily on a large-scale corpus used by online media services, including Netflix, Pinterest, and Amazon. These systems enrich recommendations by learning users' and items' embeddings projected in a low-dimensional space with two tower models (two deep neural networks), which facilitate their embedding constructs to predict users' feedback associated with items. Despite its popularity for recommendations, its theoretical behaviors remain comprehensively unexplored. We study the asymptotic behaviors of the two tower model applied in two-stage recommenders that entail a strong convergence to the optimal recommender system. We establish certain theoretical properties and statistical assurance of the two tower recommender. In addition to asymptotic behaviors, we demonstrate that recommendation with two tower architecture attains faster convergence by relying on the intrinsic dimensions of the input features. Finally, we show numerically that the two tower recommender enables encapsulating the impacts of items' and users' attributes on ratings, resulting in better performance compared to existing methods conducted using synthetic and real-world data experiments.
\end{abstract}

\section{Introduction}
Recommender systems are pivotal to enabling contents consumption for users' across various online media platforms, affecting what media items we interact, or navigate through in order to incentivize exploration. Moreover, recommender system has garnered significant attention over the past decades and have become massively popular, especially in machine learning community due to its widespread use in precision marketing and E-commerce, such as news feeding~\cite{li2016context}, movie recommendation~\cite{miller2003movielens}, online shopping~\cite{romadhony2013online}, and restaurant recommendation~\cite{vargas2011effects}. 
Content-based recommender systems~\cite{lang1995newsweeder,pazzani2007content} pertain preprocessing techniques to renovate unstructured the contents of item and the user profiles into numerical vectors. These vectors are then employed as inputs for classical machine learning algorithms, such as decision trees~\cite{middleton2004ontological}, kNN~\cite{subramaniyaswamy2017adaptive}, and SVM~\cite{oku2006context,fortuna2010real}. Collaborative filtering approaches~\cite{hofmann1999latent,schafer2007collaborative} predict a user's ratings based on the ratings of similar users or items, and employ techniques such as singular value decomposition (SVD)~\cite{mazumder2010spectral}, restricted Boltzmann machines (RBM)~\cite{salakhutdinov2007restricted}, probabilistic latent semantic analysis~\cite{hofmann2004latent}, and nearest neighbour methods~\cite{koren2009matrix}. Hybrid recommender systems~\cite{bostandjiev2012tasteweights} aim to integrate collaborative filtering and content-based filtering techniques, exemplified by the unified Boltzmann machines~\cite{gunawardana2009unified}, partial latent vector model~\cite{kouki2015hyper}, approaches utilizing user preferences~\cite{bao2022minority}, and user-item representations~\cite{lee2021bootstrapping}. A unified Boltzmann machine introduces a means of encoding both content and collaborative information as features for the purpose of rating prediction. HyPER~\cite{kouki2015hyper} presented a statistical relational learning framework capable of consolidating multi-level information sources, including user-user and item-item similarity measures, content, and social information. It uses probabilistic soft logic to make predictions, and can automatically learn to balance different information signals. The utilization of deep neural networks in recommender systems has gained widespread adoption in recent years, with various applications demonstrating significant success. Among the neural network models commonly used for recommendation systems, the two tower model~\cite{yi2019sampling,su2023beyond} has gained significant traction. This model employs two deep neural networks, known as towers, which function as encoders to embed high-dimensional features of both users and items into a low-dimensional space. The two tower model offers a significant advantage in its ability to address the well-established cold-start problem by integrating features of both users and items to generate precise recommendations for new users or items. Despite its widespread use in applications, such as book recommendation~\cite{lu2022deep}, application recommendation~\cite{yang2020mixed}, and video recommendation~\cite{yi2019sampling}, the theoretical underpinnings of the two tower model remain largely underdeveloped in the literature. This gap is particularly important given the proliferation of neural recommendation architectures such as NeuMF \cite{he2017neural}, LightGCN \cite{he2020lightgcn}, DCN-v2 \cite{wang2021dcn} that lack statistical guarantees. Our work addresses this gap by establishing asymptotic convergence rates that depend explicitly on the smoothness parameter $\beta$ and intrinsic dimension $d_{ui}$ properties that are implicitly assumed in neural methods but explicitly absent in linear baselines.\\
\textbf{Contributions:} The primary contribution of our work involves establishing asymptotic characteristics of the two tower recommender model concerning its robust convergence towards an optimal recommender system. We conduct a thorough analysis of the approximation and estimation errors of the two tower recommender model, assuming the smoothness of each embedding dimension of user or item features is a continuous function of the corresponding input characteristics. The results indicate that the robust convergence of the two tower recommender model is closely associated with the smoothness of the optimal recommender system, as well as the inherent dimensionality of the user and item features. Moreover, it is observed that the rate of convergence of the two tower model increases as the smoothness of the true model improves or the maximum intrinsic dimensions of user and item features decrease. In particular, as the underlying smoothness approaches infinity, the convergence rate of the two tower model is bounded by $O_p\left ( \left |\Omega \right |^{-1}(\log\left | \Omega \right |)^{2} \right )$, where $\Omega$ represents the set of observed ratings, and $\left | \cdot \right |$ represents the cardinality of a set. This convergence rate is faster than the majority of the existing theoretical results outlined in~\cite{zhu2016personalized}. More importantly, the established statistical guarantee for the two tower model serves as a strong theoretical justification for its successful application in a wide range of scenarios. We establish a novel theoretical connection between MSE minimization and Top-K retrieval performance (Theorem 4.6), proving that $\mathcal{R}_K(\hat{h}) \lesssim \frac{|\mathcal{I}|}{K} \cdot O_p(|\Omega|^{-\frac{2\beta}{2\beta + d_{ui}}})$ under mild margin conditions. This provides the first theoretical justification for using MSE as a surrogate objective in two-tower retrieval systems. We validate our theoretical predictions through extensive experiments on four real-world datasets showcasing that our theoretical convergence rates predict empirical performance scaling.
\section{Prior Work}
\vspace{-10pt}
\textbf{Recommendation Models:} The industry has widely embraced two-stage recommender systems, characterized by a candidate generation phase followed by a ranking process. Prominent examples of such models can be found in platforms like LinkedIn~\cite{borisyuk2016casmos}, YouTube~\cite{covington2016deep,yi2019sampling,zhao2019recommending}, and Pinterest~\cite{eksombatchai2018pixie}. Such two-stage architecture enables the real-time recommendation of highly personalized items from a vast item space. Most of such methods are dedicated to enhancing both the efficiency~\cite{yi2019sampling,kang2019candidate,chen2023fairly} and recommendation quality~\cite{chen2019top,zhao2019recommending} within the framework of this general approach, indicating a sustained commitment to refining and optimizing these models. An instance of two-stage recommendations is two tower architecture~\cite{yi2019sampling}, which represents a comprehensive framework comprising a query encoder and a candidate encoder. This architectural design has gained substantial traction in the realm of large-scale recommendation systems, as evidenced by its adoption in studies~\cite{cen2020controllable,yang2020mixed,lu2022deep}. Furthermore, it has emerged as a prominent approach in content-aware scenarios~\cite{ge2020graph}. Also, the application of two tower models within recommendation systems typically involves significantly larger corpora compared to their usage in language retrieval tasks, thereby presenting the challenge of training efficiency. Our work primarily centers on the quantification and behavioral aspects of two tower recommendation, with particular emphasis placed on optimizing the overall recommendation performance by explicitly considering the multi-level covariates information.\\
\textbf{Hybrid Recommendation Systems:} Ascertaining a singular model capable of achieving optimal performance across all scenarios is unattainable~\cite{luo2020metaselector}. 
Consequently, the simultaneous deployment of two or more recommenders is widely embraced to capitalize on their respective strengths~\cite{burke2002hybrid}. Considering that collaborative methods excel when ample data is available, while content-based recommendation exhibits superiority in cold-start situations, prior discussions have centered on a hybrid framework that combines content-based filtering with collaborative filtering. This integration facilitates a system that accommodates both new and existing users~\cite{geetha2018hybrid}. Early integration techniques typically involve computing a linear combination of individual output scores to amalgamate the outcomes produced by diverse recommenders~\cite{ekstrand2012recommenders}.
\section{Preliminaries}
\vspace{-8pt}
Let Supp($\mu$) be the spectrum (or support) of a given probability measure $\mu$. Given a function $g$ defined as $g:\mathbb{R}^{D} \rightarrow\mathbb{R}$ with its $L^{2}(\mu)$-norm and $L^{\infty}(\mu)$-norm with respect to a non-negative measure ($\mu$) are $\left \|g\right \|_{L^{2}(\mu)} = \sqrt{\int_{x}g^2(x)d\mu(x)}$ and $\left \|g\right \|_{L^{\infty}(\mu)} = \sup_{x\in \text{Supp}(\mu)}g(x)$, respectively. We denote the $l_{2}$-norm of a vector $x$ as $\left \|x\right \|_2$ which is equal to $\sqrt{\sum^{p}_{i=1} x^{2}_i}$. Given a set $S$ and an $\epsilon$-ball as the set of all points within distance $\epsilon$ of $x$ in the feature space $\mathcal{X}$, we establish the definition of $\mathcal{N}(\epsilon,S,\left \| \cdot \right \|)$ as the least number of $\epsilon$-balls required to encompass $S$ utilizing a generic metric $\left \| \cdot \right \|$.

We can define an $L$-layer neural network which can be viewed as a composition of individual functions formulated as $f(x,\Theta) = h_{L}\circ h_{L-1}\circ\ldots\circ h_{1}(x)$, where the entirety of the parameters is represented by $\Theta = ((A_{1},b_{1}),\ldots,(A_{L}, b_{L}))$, $h_{l}(x)=\sigma(A_{l}x+b_l)$ designates the $l$-th layer, and $\circ$ refers to function composition. The key components of each layer are $A_{l}\in\mathbb{R}^{p_{l}\times p_{l-1}}$ refers to the weight matrix, and $b_{l}\in\mathbb{R}^{p_{l}}$ refers to the bias term. The number of neurons in the $l$-th layer is represented by $p_l$, and $\sigma(\cdot)$ denotes an activation function that acts component-wise. Common examples of activation functions include the sigmoid function $\sigma(x) = 1/(1+\exp(-x))$ and the ReLU function $\sigma(x) = \max(x,0)$. In order to simplify notation, the expression $f(x,\Theta)$ will be represented as $f(x)$ where possible. To describe the architecture of the neural network represented by $f$, we designate the number of layers as $U(f)$, the parameter scale is defined as the maximum value of the infinity norm of the bias vector $b_l$ and the vectorized weight matrix $A_l$, taken over all layers $l$ in $f$. It is represented by $D(f)=\max_{l=1,\ldots,U(f)}\max\left \{\left \|b_{l} \right \|_{\infty}, \left \|\text{vec}(A_l) \right \|_{\infty} \right \}$ and the number of effective parameters as $Z(f)=\sum_{l=1}^{U(f)}(\left |b_l \right |_0 + \left |\text{vec}(A_l) \right |_0)$, where $\text{vec}(\cdot)$ is a function that converts a matrix into a vector. Now, we leverage the concept of H\"{o}lder space\footnote{https://en.wikipedia.org/wiki/Hölder\_condition}, which is a space of functions that are defined on a given domain and satisfy certain conditions related to their smoothness and regularity~\cite{chen2019efficient}. Specifically, we can define a function space of H\"{o}lder continuous functions and use it to approximate the unknown user-item preference function~\cite{liu2021smooth}. The degree of smoothness or regularity of the function can be controlled by choosing an appropriate value of the H\"{o}lder exponent. Assuming a degree of smoothness $\beta\geq 0$, the H\"{o}lder space can be defined as follows $\mathcal{H}(\beta, \left [0,1\right]^D) = \{f\in C^{\left \lfloor\beta\right \rfloor}(\left [0,1 \right]^D)\mid\left \|f \right \|_{\mathcal{H}(\beta, \left [0,1\right]^D)}<\infty \}$, here, the set $C^{\left \lfloor\beta\right \rfloor}(\left [0,1 \right]^D)$ comprises all functions that have ${\left \lfloor\beta\right \rfloor}$ times differentiable and continuous derivatives on the domain $\left [0,1\right]^D$, where $\left \lfloor\cdot\right \rfloor$ represents the floor function. The H\"{o}lder norm is described as follows,
\begin{eqnarray*}
    \left \|f \right \|_{\mathcal{H}(\beta, \left [0,1\right]^D)} = \max_{\alpha:\left \|\alpha \right \|_1<\beta} \sup_{x\in\left [0,1 \right]^D}\left |\partial^\alpha f(x) \right|+
    \max_{\alpha:\left \|\alpha \right \|_1=\left \lfloor\beta\right \rfloor} \sup_{x,x^{'}\in\left [0,1 \right]^D, x\neq x^{'}}\frac{\left |\partial^\alpha f(x)-\partial^\alpha f(x^{'}) \right|}{\left \|x-x^{'} \right \|^{\beta-\left \lfloor\beta\right \rfloor}_{\infty}}
\end{eqnarray*}
where the H\"{o}lder exponent~$\alpha_i \geq 0$ is an integer with $\alpha = (\alpha_1,\ldots,\alpha_D)$, and $\partial^\alpha f=\partial^{\alpha_1}_1,\ldots,\partial^{\alpha_D}_D$. Additionally, we generalize the H\"{o}lder space $\mathcal{H}(\beta, \left [0,1\right]^D, M)=\left \{f\in\mathcal{H}(\beta, \left [0,1\right]^D)\mid \left \|f \right \|_{\mathcal{H}(\beta, \left [0,1\right]^D)}\leq M \right \}$ to be considered as a closed ball with radius $M$, and so $\mathcal{H}^{p}(\beta, \left [0,1\right]^D, M) = \mathcal{H}(\beta, \left [0,1\right]^D, M)\times \mathcal{H}(\beta, \left [0,1\right]^D, M)\times\ldots\times\mathcal{H}(\beta, \left [0,1\right]^D, M)$.

\subsection{Two Tower Recommendation Model}
\begin{figure}[!htbp]
    \centering
    \begin{minipage}{0.35\textwidth}
        \centering
        \includegraphics[width=\linewidth]{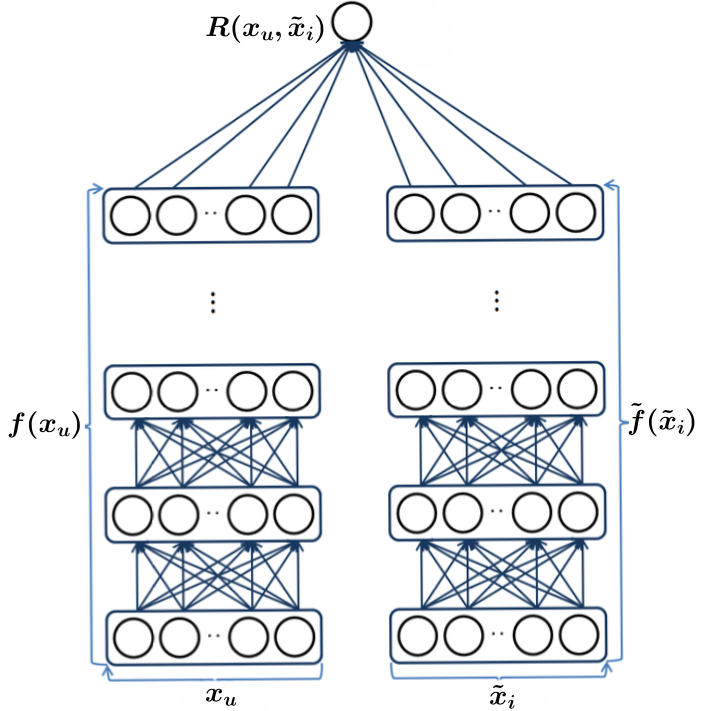}
        \caption{\textbf{An illustration of the Two Tower recommender model.}}
        \label{fig:t2rec}
    \end{minipage}\hfill
    \begin{minipage}{0.6\textwidth}
        \begin{equation}\label{ttopt}
            R(x_u,\tilde{x}_i) = \langle f(x_{u}), \tilde{f}(\tilde{x}_{i}) \rangle 
        \end{equation}
        \begin{align}\label{cost}
            \min_{f,\tilde{f}}\frac{1}{\left |\Omega \right |}\sum_{(u,i)\in\Omega}(k_{ui}- \langle f(x_{u}), \tilde{f}(\tilde{x}_{i}) \rangle)^2 + \lambda \left \{J(f)+J(\tilde{f}) \right \}
        \end{align}
    \end{minipage}
\end{figure}

Our focus in this work is on a specific recommender system approach, namely the two tower model~\cite{yi2019sampling}, which is a neural network architecture that is often used for two-stage recommendation. The two towers of the model are responsible for the candidate retrieval and ranking stages, respectively. Two tower models are capable of learning complex relationships between users and items, and it is able to scale to large datasets. In numerous recommender models, covariates are unstructured and high-dimensional, and may include information such as user profiles and textual item descriptions. There is a prevailing belief that such information can often be represented in a low-dimensional intrinsic form, and can be seamlessly incorporated into the feature engineering phase of a deep learning model. For a standard recommender model with user covariates denoted as $x_{u}\in\mathbb{R}^{D_u}$ and item covariates as $\tilde{x}_{i}\in\mathbb{R}^{D_i}$, the two tower model is formulated as in given Equation~\ref{ttopt} and a schematic overview in Figure~\ref{fig:t2rec}. The two deep neural networks are described as $f:\mathbb{R}^{D_u}\rightarrow\mathbb{R}^p$ and $\tilde{f}:\mathbb{R}^{D_i}\rightarrow\mathbb{R}^p$ delineating $x_u$ and $\tilde{x}_i$ into the same $p$-dimensional embedding space. The two tower model follows recommendation approach to be based on the dot product between the feature vectors extracted from the two towers, $f(x_{u})$ and $\tilde{f}(\tilde{x}_{i})$. The cost function to optimize the two tower model can be structured via Equation~\ref{cost}.
This given equation represents the cost function of the two tower model, where $J(\cdot)$ can be the penalty term of $L_1$-norm or $L_2$-norm to avoid overfitting in the deep neural network, and $k_{ui}$ refers to the number of layers in the user/item towers. The optimization problem presented in Equation~\ref{ttopt} can be efficiently solved using an established open-source neural network library such as PyTorch~\cite{paszke2019pytorch}. We detail this in section~\ref{prem:append}. Given that when the user and item covariates $x_u$ and $\tilde{x}_i$ respectively are encoded using one-hot encoding, the Equation~\ref{ttopt} simplifies to the conventional collaborative filtering approach based on SVD. The two tower model represents a hybrid recommendation which combines the advantages of collaborative filtering and content-based filtering methods by utilizing low-dimensional representations for both users and items. The use of deep neural network structure facilitates the flexible representation of users and items, and enables the capture of non-linear covariate effects, which is not possible with linear modeling~\cite{bi2017group,mao2019matrix}. Subsequently, the two tower model can mitigate the cold-start problem by incorporating new users and items using their respective covariate representations~\cite{van2013deep}. We highlight that the optimization task presented in Equation~\ref{cost} offers a general framework for developing deep recommender systems, and the two neural network structures can be modified to suit various data sources, such as using for the sequential data~\cite{twardowski2016modelling} via a recurrent neural network (RNN) or for images via a convolutional neural network (CNN)~\cite{truong2019multimodal,yu2019visual}.

\section{Asymptotic Behaviors}
\vspace{-8pt}
We aim to establish some theoretical properties of the two tower model, which relate to its strong convergence to the true model. This is considered one of the initial efforts in quantifying the asymptotic behaviors of deep recommender systems. Our work focuses on examining the model's properties and establishing a theoretical foundation to support its reliability and effectiveness in producing accurate recommendations. Assuming that the given model generates the observed data $\{(x_u,\tilde{x}_i,k_{ui}), (u,i)\in\Omega \}$, provided $\tilde{x}_i \in\left [0,1\right]^{D_i}$, $x_u \in\left [0,1\right]^{D_u}$, and $\epsilon_{ui}$ comprise a sub-Gaussian noise bounded by $B_e$ with variance $\sigma^{2}$, which are independent and identically distributed. Also, it follows based on the H\"{o}lder norm that $f^{\ast}=(f^{\ast}_1,\ldots,f^{\ast}_p)$, $\tilde{f}^{\ast} = (\tilde{f}^{\ast}_1,\ldots,\tilde{f}^{\ast}_p)$ with $f^{\ast}_j \in \mathcal{H}(\beta, \left [0,1\right]^{D_u}, M)$ and $\tilde{f}^{\ast}_j \in \mathcal{H}(\beta, \left [0,1\right]^{D_i}, M)$, where $\sup_{x\in\left [0,1\right]^{D_u}} \left |f^{\ast}_j(x) \right |\leq M$ and $\sup_{x\in\left [0,1\right]^{D_i}} \left |\tilde{f}^{\ast}_j(x) \right |\leq M$ for all $j=1,\ldots,p$.
\begin{align}\label{model}
    k_{ui} = R^{\ast}(x_u,\tilde{x}_i)+\epsilon_{ui}=\langle f^{\ast}(x_{u}),\tilde{f^{\ast}}(\tilde{x}_{i}) \rangle+\epsilon_{ui},
\end{align}

\subsection{Problem Formulation and Analysis}
We characterize two classes of deep neural network with bounded parameters for users and items as $\mathcal{F}_{D_{u}}(W,L,B,M) = \{f\mid Z(f)\leq W, U(f)\leq L, D(f)\leq B,\sup_{x\in\left [0,1 \right]^{D_u}} \max_{j=1,\ldots,p} \left |f_j(x) \right |\leq 2M \},\mathcal{F}_{D_{i}}(\tilde{W},\tilde{L},\tilde{B},M)= \{\tilde{f}\mid Z(\tilde{f}) \leq \tilde{W}, U(\tilde{f}) \leq \tilde{L},\\ D(\tilde{f})\leq \tilde{B},\sup_{x\in\left [0,1 \right]^{D_i}}\max_{j=1,\ldots,p}\left |\tilde{f}_{j}(x) \right |\leq 2M \},$
provided assumption of the boundedness of $f$ and $\tilde{f}$ is made in order to reduce the dimensionality of the parameter space required for the approximation. For the sake of conciseness, we designate $\mathcal{F}_{D_{u}}(W,L,B,M)$ and $\mathcal{F}_{D_{i}}(\tilde{W},\tilde{L},\tilde{B},M)$ as $\mathcal{F}_{D_{u}}$ and $\mathcal{F}_{D_{i}}$. Additionally, we further provide a definition for the class of deep recommender systems as $\mathcal{R}^{\Phi}= \{R(x_u,\tilde{x}_i)=\langle f(x_{u}),\tilde{f}(\tilde{x}_{i}) \rangle\mid f\in\mathcal{F}_{D_{u}} (W,L,B,M), \tilde{f}\in\mathcal{F}_{D_{i}}(\tilde{W},\tilde{L},\tilde{B},M) \}$, where $\Phi = (W,L,B,M,\tilde{W},\tilde{L},\tilde{B})$ represents the parameters required for estimating the size of $\mathcal{R}^{\Phi}$. The estimate $\hat{R}$ can be formulated as 
\begin{align*}
    \hat{R}=\arg\min_{R\in\mathcal{R}^{\Phi}}\frac{1}{\left |\Omega \right |}\sum_{(u,i)\in\Omega}(k_{ui}-R(x_u,\tilde{x}_i))^2 + \lambda \underbrace{\{\sum_{l=1}^{L}(\left \|A_l \right \|^{2}_F + \left \|b_l \right \|^{2}_2)+\sum_{l=1}^{\tilde{L}}(\left \|\tilde{A}_l \right \|^{2}_F + \left \|\tilde{b}_l \right \|^{2}_2) \}}_{J(R)}
\end{align*}
Our main formulation in this section is to estimate the approximation error of the two tower model, which we define as a deep recommender system. Our proposed strategy in Theorem~\ref{th1}, which presents and incorporates the existing theoretical approaches in~\cite{nakada2020adaptive} but has been adapted to consider specific challenges faced in deep recommender systems. One of these challenges is the high-dimensionality of the input for these systems, which often reside on a low-dimensional manifold, particularly in cases where the input data contains sparse binarized features such as one-hot encoding or bag-of-words. To estimate the intrinsic dimension of the input space $S$, we define its upper Minkowski dimension based on~\cite{falconer2004fractal} as $\dim(S)= \inf\left \{d^{\ast}\geq 0 \mid\lim_{\epsilon\rightarrow 0}\sup\mathcal{N}(\epsilon,S,\left \|\cdot\right \|_\infty)\epsilon^{d^{\ast}}=0 \right \}$. It is worth noting that the upper Minkowski dimension of a discrete input space is invariably 0. Therefore, when binarized features are included in the input of deep recommender systems, they typically do not enhance the upper Minkowski dimension.
\begin{theorem}\label{th1}
Let $\dim(\text{Supp}(\mu_u))\leq d_u$ $\dim(\text{Supp}(\mu_i))\leq d_i$ be the given Minkowski dimension, provided the probability measure of $x_u$ and $\tilde{x}_i$ refers to $\mu_u$ and $\mu_i$, respectively. Then, for any $\epsilon >0$, $\exists~\Phi = (W,L,B,M,\tilde{W},\tilde{L},\tilde{B})$ with $W=O(\epsilon^{-d_u/\beta}), \tilde{W}=O(\epsilon^{-d_i/\beta})$, and $B=O(\epsilon^{-s})$, and $\tilde{B}=O(\epsilon^{-s})$, such that 
\begin{align*}
    \inf_{R\in\mathcal{R}^{\Phi}}\left \|R-R^{\ast} \right \|_{L^{\infty}(\mu_{ui})}\leq 3pM\epsilon
\end{align*}
where $\mu_{ui}$ represents the probability measure of $(x_u,\tilde{x}_i)$ on $\text{Supp}(\mu_u)\times\text{Supp}(\mu_i)$.
\end{theorem}
Theorem~\ref{th1} provides a measure of the approximation error of the two tower model. The upper bound on the approximation error in Theorem~\ref{th1} implies that there exists some $\Phi$ for which the true model can be effectively approximated by $\mathcal{R}^{\Phi}$, provided that the underlying true functions $f^{\ast}$ and $\tilde{f}^{\ast}$ in Equation~\ref{model} have sufficient smoothness. Moreover, Theorem~\ref{th1} remains valid irrespective of the value of L, indicating that the approximation error of the two tower model can converge to zero with any number of layers. We defer the proof to the appendix~\ref{append}.

\subsection{Robust Convergence}
In order to establish the robust convergence of the two tower model, introductory lemmas are required to quantify its entropy. These lemmas are essential for measuring the estimation error of $\hat{R}$ and balancing it with the approximation error. Therefore, we propose the following lemmas to evaluate the entropy of $\mathcal{R}^{\Phi}$, which is a critical factor in deriving the estimation error of $\hat{R}$.
\begin{lemma}\label{lem:l1}
Let the functional space be $\mathcal{K}_D (W,L,B,M)=
\{f(\cdot;\Theta):\Theta\in S_{B}(2W,D)
\times S^{L-2}_{B}(2W,2W) 
\times S_{B}(p,2W) \},$
where $S_{B}(c,d)= \{(A,b)\mid A\in\left [-B,B \right]^{c\times d},b\in\left [-B,B \right]^{c} \}$. There exists a mapping $Q:\mathcal{F}_D (W,L,B,M)\rightarrow\mathcal{K}_D (W,L,B,M)$ such that $f(x)=Q(f)(x)$ for any $f\in\mathcal{F}_D (W,L,B,M)$ provided $Z(Q(f))\leq 14LW\log W$.
\end{lemma}
The functional spaces $\mathcal{F}_{D_u}$ and $\mathcal{F}_{D_i}$ comprise neural networks with distinct layer architectures and widths, rendering it difficult to establish their entropy in a manner that is amenable to analysis. However, Lemma~\ref{lem:l1} establishes that $\mathcal{F}_{D_u}$ and $\mathcal{F}_{D_i}$ can be embedded into larger functional spaces $\mathcal{K}_{D_u}$ and $\mathcal{K}_{D_i}$ that consist of deep neural networks with consistent dimensions. Consequently, the entropy of $\mathcal{K}_{D_u}$ and $\mathcal{K}_{D_i}$ can be directly estimated as a parametric model~\cite{zhang2002covering}, thereby providing an upper bound for the entropy of $\mathcal{F}_{D_u}$ and $\mathcal{F}_{D_i}$, respectively. Furthermore, it is crucial to note that the effective number of parameters in $\mathcal{K}_{D}$ is of the same magnitude as that of $\mathcal{F}_{D}$, except for a trivial logarithmic term.
\begin{lemma}\label{lem:l2}
For any $f(x;\Theta),f'(x;\Theta')\in\mathcal{K}_D (W,L,B,M)$, it remains valid that
$\sup_{\left \|x \right \|_{\infty}\leq 1}\left \| f(x;\Theta)-f'(x;\Theta') \right \|_2 \leq pC(W,L,B)\epsilon, ~\text{given~that}~ \| \Theta-\Theta' \|_{\infty}$
 $\leq\epsilon$~\text{where}~ $C(W,L,B)=(WB)^{L}\left (\frac{L}{B}+\frac{L}{WB-1} \right)-\left (\frac{(WB)^{L}-1}{(WB-1)^2} \right )$.  
\end{lemma}
Lemma~\ref{lem:l2} introduces a continuity property of H\"{o}lder-type for the neural networks that belong to $\mathcal{K}_D(W,L,B,M)$. Here, the term $C(W,L,B)$ may tend to infinity concerning the dimensions W, L, and B. This continuity property enables the computation of the functional class's entropy for the neural networks associated with users and items in the following Lemma~\ref{lem:l3}, the proof of which is deferred to the appendix~\ref{append}.

\begin{lemma}\label{lem:l3}
Given $\Phi$=$(W,L,B,M,\tilde{W},\tilde{L},\tilde{B})$, it remains valid that, 
$\log\mathcal{N}_{\left[\cdot \right]}(\epsilon,\mathcal{R}^{\Phi},\left \|\cdot \|_{L^{2}(\mu_{ui})} \right)\leq C_2(W\log{W}+\tilde{W} 
\log{\tilde{W}})\log (\epsilon^{-1} C_3(C(W,L,B)+C(\tilde{W},\tilde{L},\tilde{B})))$,
provided $\mathcal{N}_{\left[\cdot \right]}(\epsilon,\mathcal{R}^{\Phi},\left \|\cdot \right \|_{L^{2}(\mu_{ui})})$ depicts the $\epsilon$-bracketing quantity of $\mathcal{R}^{\Phi}$ with respect to the metric $\left \|\cdot \right \|_{L^{2}(\mu_{ui})}$, $C_2 = 28\max \{L, \tilde{L} \}$, $C_3 = 2p^{3/2}M\max \{B, \tilde{B} \}$, and $C(\cdot,\cdot,\cdot)$ is stipulated as per Lemma~\ref{lem:l2}.
\end{lemma}
Lemma~\ref{lem:l3} ascertains an upper limit on the bracketing entropy of the two tower recommender model, thereby serving as a fundamental component in deducing the estimation error associated with the two tower recommender model. This inference is accomplished through the application of empirical process theory and certain large deviation inequalities. The use of identical measures of entropy has also been employed in seminal work~\cite{zhou2002covering} to quantify the expressive capacity of diverse functional classes.

\begin{theorem}\label{thm2}
If all the conditions described in Theorem~\ref{th1} are realized, then it remains valid that 
\begin{multline*}
P(\|\hat{R}-R^{\ast} \|_{L^2 (\mu_{ui})}^{2}\leq L_{ui}\left |\Omega \right |^{-2\beta/(2\beta+d_{ui})}(\log\left |\Omega \right |)^2 )\geq 1-24\exp(-C_1 \left |\Omega \right |^{d_{ui}/(2\beta+d_{ui})} \log\left |\Omega \right|)\\
\text{given}~4\lambda_{\left |\Omega \right |}J(R_0)\leq L_{ui}\left |\Omega \right|^{-2\beta/(2\beta+d_{ui})}\log\left |\Omega \right |, \text{where},~ L_{ui} =\max\{L, \tilde{L} \}~ \text{with}~ \\L=O(\beta\log_2\beta/d_{u})~ \text{and}~ \tilde{L}=O(\beta\log_2\beta/d_{i}),
C_1=6\max\{(50p^2M^4+4\sigma^2),1\}(25p^2M^4+B_e^2)/13, \\ B_e=O(\left |\Omega \right |^c)~ \text{for}~ c< d_{ui}/(4\beta+2d_{ui})~ \text{in which}~ d_{ui}=\max\{d_u,d_i\}.
\end{multline*} 
The underlying parameters of~ $\mathcal{R}^{\Phi}$ are $W$ and $\tilde{W}$ which equals
$O(\left |\Omega \right|^{d_{ui}/(2\beta+d_{ui})}\log\left |\Omega \right |), B=O(\left |\Omega \right|^{2\beta s/(2\beta+d_{u})}\log\left |\Omega \right |), \text{and}~ \tilde{B}=O(\left |\Omega \right|^{2\beta s/(2\beta+d_{i})}\log\left |\Omega \right |)$.
\end{theorem}
Theorem~\ref{thm2} provides evidence of the convergence of the two tower recommender model towards the true model at a rapid rate, explicitly determined by the values of $\beta$, $d_u$, and $d_i$. 
Here $B_e$ denotes the sub-Gaussian norm of the noise distribution controlling the tail behavior of the rating noise $\epsilon_{ui}$. The condition $c$ ensures that the noise does not dominate the signal in the asymptotic regime. In particular, when $\beta$ attains a competent magnitude, the convergence rate approximates $O_p(\left |\Omega \right|^{-1}(\log\left |\Omega \right |)^2)$, surpassing the majority of existing findings~\cite{zhu2016personalized}. This advantage arises primarily from the smooth representation of covariates provided by the latent embeddings of users and items, resulting in a significantly reduced number of parameters compared to conventional collaborative filtering approaches. As a consequence, the two tower recommender model exhibits an accelerated rate of convergence. Furthermore, it is intriguing to observe that with predefined $\beta$, $d_u$, and $d_i$, the value of $L_{ui}$ remains constant. This implies that finite depths of the two tower recommender model are adequate for approximating the true model, while the widths of the user network and item network increase at a rate of $O(\left |\Omega \right|^{d_{ui}/(2\beta+d_{ui})}\log\left |\Omega \right |)$. A detailed proof of which is deferred to the appendix~\ref{append}.
\subsubsection{Theoretical Guarantees for Top-k Retrieval Objectives}
\begin{theorem}\label{thm3}
Let $\mathcal{R}_{K}(h)$ denote the expected Top-$k$ retrieval error (mis-retrieval risk) for a two-tower model $h(\mathbf{x}_u, \tilde{\mathbf{x}}_i) = \langle f(\mathbf{x}_u), \tilde{f}(\tilde{\mathbf{x}}_i) \rangle$, defined as the probability that the ground-truth relevant item $i^*$ is ranked below the $k$-th position among a set of items $\mathcal{I}$:$$\mathcal{R}_{K}(h) = \mathbb{E}_{(u, i^*) \sim \mathcal{D}} \left[ \mathbb{I}( \text{rank}(i^* | u, h) > K ) \right]$$Under the assumptions of Theorem 4.5, and assuming the ground-truth score gap between relevant item $i^*$ and any irrelevant item $j$ satisfies a margin condition $\Delta = R^*(u, i^*) - R^*(u, j) > 0$, the Top-$k$ retrieval error converges as:$$\mathcal{R}_{K}(\hat{h}) \lesssim \frac{|\mathcal{I}|}{K} \cdot O_p\left(|\Omega|^{-\frac{\beta}{2\beta + d_{ui}}}\right)$$where $|\mathcal{I}|$ is the total item corpus size. This guarantees that as sample size $|\Omega|$ increases, the probability of missing the relevant item in the top-$k$ set vanishes, confirming the model's effectiveness as a candidate generator.
\end{theorem}
Theorem~\ref{thm3} bridges the gap between the pairwise error and the Top-k classification metric. It informs that optimizing the two tower model minimizes the Top-$k$ retrieval error, validating its use for recall-oriented tasks. The error scales linearly with the corpus size $|\mathcal{I}|$ and inversely with $K$, providing a justification for using larger $K$ in the retrieval stage to ensure recall. This bound assures that as the estimator $\hat{h}$ converges to $R^*$, the probability of missing the relevant item in the retrieval set vanishes. The term $\frac{|\mathcal{I}|}{K}$ explicitly captures the retrieval nature, a larger candidate set size $K$ linearly reduces the risk of a miss. A detailed proof of this is reported in the appendix~\ref{append}. We also provide a theoretical lens on system constraints introduced as a constrained optimization problem in Theorem~\ref{thm4}, where the constraint (latency) directly influences achievable performance by limiting the richness of the function class.

\section{Experiments}\label{sec:exp}
We present a thorough numerical evaluation of the two tower recommender model, represented as T$^2$Rec is conducted on various synthetic and real-world datasets. We compare its performance against a range of established competitors, including regularized SVD (rSVD), SVD++, co-clustering algorithm (Co-Ca), and K-nearest neighbors (KNN), and state-of-the-art neural baselines NeuMF \cite{he2017neural}, LightGCN \cite{he2020lightgcn}, and DCN-v2 \cite{wang2021dcn}. We implement T$^2$Rec framework via TensorFlow~\citep{tensorf},
while the other baseline models' implementations are accessible in the Pythonic library\footnote{https://surpriselib.com} of simple recommendation system engine~\citep{hug2020surprise}. The rSVD method uses an alternative least square (ALS) algorithm for estimating latent factors of users and items. SVD++ utilizes stochastic gradient descent (SGD) to minimize a regularized squared error objective. Co-Ca categorizes users and items into clusters that are assigned distinct baseline ratings. SlopeOne is primarily an item-based collaborative filtering approach that leverages ratings of similar items for prediction, and KNN predominantly exploits the weighted average of the ratings of the top-K most similar users for prediction. For neural baselines, NeuMF combines generalized matrix factorization and MLP, LightGCN simplifies graph convolutional networks by removing feature transformation and nonlinear activation, and DCN-V2 uses cross networks to learn explicit feature interactions. We also validate the theoretical claims established in Theorems 4.1, 4.5, and 4.6 by systematically varying the key parameters such as network width $W$, depth $L$, intrinsic dimension $d_{ui}$, H\"{o}lder smoothness $\beta$, and sample size $|\Omega|$.
\\
\textbf{Training Settings:} The present study involves tuning parameters for several methods through grid search. To accomplish this, the datasets are partitioned into two sets, one for training and the other for testing. For the training set, the optimal model for SVD++, KNN, rSVD, and Co-Ca is selected based on 5-fold cross-validation of the training set. Meanwhile, the optimal model for T$^2$Rec and neural baselines is determined using a validation set that is 20\% of the size of the training set. This approach helps to reduce the computational cost associated with cross-validation. The hyperparameters related to the regularization parameter $\lambda$ in T$^2$Rec and rSVD are defined as grid values $10^{-6+k/3}$, where k=$\{0,1,\ldots,24\}$. The hyperparameters for the number of clusters in Co-Ca and the neighborhood parameter K in KNN are determined by specifying a grid of possible values $\{5,10,\ldots,50\}$. The KNN algorithm employs a similarity measure based on the mean square similarity difference of common ratings between any two users or items~\citep{hug2020surprise}. For T$^2$Rec and neural baselines, which is a deep neural network-based method, the SGD learning rate is initialized to $1\mathrm{e}{-2}$ and has a decay rate of 0.9 and a minimum learning rate of $5\mathrm{e}{-3}$. To prevent overfitting, an early-stopping scheme is utilized.
\subsection{Results on Synthetic Instances}
We investigate different scenarios of a synthetic example, wherein we set the sizes of the rating matrix $R_m=\{r_{ui}\}_{1\leq u\leq n, 1\leq i\leq m}$ as $(n,m)$ = (1500,1500), (2000,2000), and (3000,3000), while keeping the number of observed ratings fixed at 100k. This leads to sparsity levels ranging from 0.011 to 0.044. Secondly, we define the nominal dimensions of $x_u$ and $\tilde{x}_i$ represented by $D_u$ and $D_i$ is 50. The dimension of the representation $p$ is 30 and the users and items true functions is formulated as 
\begin{gather*}
    f^{\ast}(x_u) = (f^{\ast}_{1}(x_u),\ldots, f^{\ast}_{p}(x_u)), \text{and}~ \tilde{f}^{\ast}(\tilde{x}_u) 
    =(\tilde{f}^{\ast}_{1}(\tilde{x}_u),\ldots, \tilde{f}^{\ast}_{p}(\tilde{x}_u)). \\
\text{Given}~ f^{\ast}_{j}(x_u) = \sum_{l=1}^{D_u}\alpha_{jl}\sin(2\pi x_{ul})+\sum_{l=1}^{D_u}\beta_{jl}\cos(2\pi x_{ul})+\sum_{l=1}^{D_{u}-1}\zeta_{jl}x_{ul}x_{u(l+1)}~\text{and}~\\ \tilde{f}^{\ast}_{j}(\tilde{x}_u) = \sum_{l=1}^{D_i}\tilde{\alpha}_{jl}\sin(2\pi \tilde{x}_{il})+\sum_{l=1}^{D_i}\tilde{\beta}_{jl}\cos(2\pi\tilde{x}_{il})+\sum_{l=1}^{D_{i}-1}\tilde{\zeta}_{jl}\tilde{x}_{il}\tilde{x}_{i(l+1)},
\end{gather*}
$\text{where}~ \alpha_{jl},\tilde{\alpha}_{jl}, \beta_{jl}, \tilde{\beta}_{jl}, \zeta_{jl}, \text{and}~ \tilde{\zeta}_{jl}$ drawn uniformly from a sample region of $[-0.15,0.15]$. To replicate the low inherent dimensionality of covariates, we sample $x_{ul}$ and $\tilde{x}_{il}$ with $l=1,\ldots,d$, from $[0,1]$, and it can be updated as $x_{ul}=x_{u(l-d)}$ and $\tilde{x}_{il}=\tilde{x}_{i(l-d)}$, provided $l=d+1,\ldots,50$ and the intrinsic dimension $d\in\{20,30,40\}$. Ultimately, the ratings are produced by the subsequent model, $r_{ui}=\left \langle f^{\ast}(x_u),\tilde{f}^{\ast}(\tilde{x}_u) \right \rangle + \epsilon_{ui}$, where $\epsilon_{ui}$ delineates a Gaussian distribution with  the mean of zero and variance of 0.1. For each case, the deep neural networks are configured for both users and items in the two tower recommender model (T$^2$Rec) as a five-layer fully-connected neural network comprising 50 neurons in each hidden layer and 30 neurons in the output layer. The root mean square errors (RMSE) are calculated and averaged across each baseline models, and their standard error(s) (SE) are also computed which is reported in Table~\ref{tab:synthetic}.
\begin{table*}[btp]
\caption{\label{tab:synthetic} Performance comparison of varied baseline models along with our proposed approach for T$^2$Rec is reported. The size of the rating matrix size$(K)$ and d represents the intrinsic dimension. The RMSE values are averaged over 50 replications and each of the model variants report the standard errors. Our proposed approach for T$^2$Rec stands best in each scenario as shown in the shaded part.}
   \centering 
   \resizebox{\textwidth}{!}{
   \begin{tabular}{|c|c|c|c|c|c|c|c|c|c|c|}
    \hline 
       \multirow{1}{*}{\textbf{\diagbox{size(K),d}{Model}}} &
       \multicolumn{2}{c|}{\textbf{rSVD}} & 
       \multicolumn{2}{c|}{\textbf{KNN}} &
       \multicolumn{2}{c|}{\textbf{Co-Ca}} &
       \multicolumn{2}{c|}{\textbf{SVD++}} &
       \multicolumn{2}{c|}{\textbf{T$^2$Rec}} \\
      & RMSE & SE & RMSE & SE & RMSE & SE & RMSE & SE & RMSE & SE\\
      \hline
      (1500,1500),20 & 1.566 & 0.008 & 1.990 & 0.013 & 1.815 & 0.012 & 1.507 & 0.008 & \ccw{0.496} & \ccw{0.011} \\
      \hline
      (1500,1500),30 & 1.742 & 0.009 & 2.063 & 0.011 & 1.944 & 0.010 & 1.704 & 0.007 & \ccw{1.330} & \ccw{0.010} \\
      \hline
      (1500,1500),40 & 1.845 & 0.007 & 2.075 & 0.012 & 2.015 & 0.011 & 1.806 & 0.007 & \ccw{1.604} & \ccw{0.01} \\
      \hline
      (2000,2000),20 & 1.908 & 0.010 & 2.074 & 0.016 & 1.907 & 0.013 & 1.849 & 0.008 &\ccw{0.438} & \ccw{0.022} \\
      \hline
      (2000,2000),30  & 2.041 & 0.013 & 2.120 & 0.012 & 2.027 & 0.013 & 1.995 & 0.011 & \ccw{1.358} & \ccw{0.013} \\
      \hline
      (2000,2000),40  & 2.110 & 0.010 & 2.150 & 0.010 & 2.089 & 0.010 & 2.073 & 0.009 & \ccw{1.703} & \ccw{0.009} \\
      \hline
      (3000,3000),20 & 2.105 & 0.023 & 2.301 & 0.024 & 2.149 & 0.022 & 2.198 & 0.021 & \ccw{0.373} & \ccw{0.010} \\
      \hline
      (3000,3000),30 & 2.196 & 0.012 & 2.311 & 0.012 & 2.246 & 0.015 & 2.204 & 0.012 & \ccw{1.353} & \ccw{0.013} \\
      \hline
      (3000,3000),40 & 2.209 & 0.020 & 2.338 & 0.021 & 2.291 & 0.021 & 2.219 & 0.019 & \ccw{1.862} & \ccw{0.010} \\
      \bottomrule
      \multicolumn{11}{c}{\textbf{Results on Yelp dataset}} \\
      \toprule
       Cold-start & 1.058 & 0.0002 & 1.058 & 0.0002 & 1.058 & 0.0002 & 1.058 & 0.0002 & \ccg{0.965} & 0.0004 \\
       \hline
       Warm-start & 0.965 & 0.0007 & 1.045 & 0.0008 & 1.055 & 0.0008 & \ccg{0.907} & 0.0006 & \ccg{0.955} & 0.0007 \\
       \hline
       Overall & 1.032 & 0.004 & 1.054 & 0.0004 & 1.057 & 0.0004 & 1.037 & 0.0003 & \ccg{0.962} & 0.0004 \\
       \hline
   \end{tabular}}
\end{table*}

\subsection{Results on a Real-World Dataset}
We employ an open source dataset of Yelp\footnote{https://www.yelp.com/dataset}, which comprises four varied linked components such as user, review, business, and check-in. The user segment entails personal information for approximately 5.2M Yelp community users, encompassing their number of reviews, count of fans, elite experience status, and personal social network information. Furthermore, behavioral aspects of users, such as the average rating given to reviews and voting data received from other users such as `useful', `funny', and `cool', are also included. The business segment offers information about the location, latitude-longitude coordinates, review counts, and categories for nearly 174k businesses. Within the review segment, each review encompasses information about the user, the associated business, the textual comment, and the corresponding star rating for that business. Finally, the `check-in' segment provides the counts of check-ins recorded at each business. By leveraging the user, business, and review segments, we generate profiles for users and businesses, which will subsequently serve as covariates in the T$^2$Rec model analysis. For data preprocessing, we rely on cities that contain a minimum of 20 businesses, given businesses that have accumulated at least 100 reviews. This selection criteria yields a set of 15,090 businesses in the item set. For each business, we employ one-hot encoding to numerically represent their `location' and `category', utilizing these variables as part of the item covariates. In terms of users, we gather information on their elite experience, which is binary in nature and indicates whether they have ever held elite user status within the Yelp community. Additionally, we consider the overall feedback they have received, including ratings such as `useful', `cool', and `funny'. A comprehensive results on real datasets are reported in Section~\ref{realdata} including Table~\ref{tab:experimental-results2}.
\subsection{Ablation and Theoretical Analysis}\vspace{-8pt}
We perform synthetic ablation studies to isolate the impact of intrinsic dimension ($d_{ui}$) and smoothness ($\beta$) on the convergence rate, controlling for all other factors. To validate the convergence bound derived in Theorem 4.5, $\|\hat{R} - R^*\|^2 \lesssim |\Omega|^{-\frac{2\beta}{2\beta + d_{ui}}}$, we implemented a controlled synthetic environment by generating user and item features on a low-dimensional latent manifold $\mathcal{Z} \subseteq \mathbb{R}^{d_{ui}}$, which were then mapped non-linearly to a high-dimensional nominal space $\mathbb{R}^{D}$ ($D=128$) using sinusoidal embeddings to simulate complex manifold structures. The true preference function is defined as $R^*(\mathbf{z}_u, \mathbf{z}_i) = \tanh(\langle \mathbf{z}_u, \mathbf{z}_i \rangle)^\gamma$. The exponent $\gamma$ serves as a proxy for the inverse of smoothness, where $\gamma=1.0$ represents a standard smooth interaction (high $\beta$), while $\gamma=2.0$ induces sharper transitions (lower $\beta$). We trained a standard two tower deep neural network with hidden dimension 64, embedding dimension 32 using the Adam optimizer across a log-spaced sweep of sample sizes $|\Omega| \in [2\times 10^3, 10^5]$. We observe a distinct power-law relationship between the sample size $|\Omega|$ and the MSE. The slope of the linear fit in the log-log scale plot reported in Fig.~\ref{fig:intd} corresponds directly to the convergence exponent $-\frac{2\beta}{2\beta + d_{ui}}$. With $d_{ui}=2$, the model achieves the steepest convergence slope, indicating high sample efficiency. And $d_{ui}=8$ indicates that the slope significantly flattens (becomes less negative). Quantitatively, moving from $d_{ui}=2$ to $d_{ui}=8$ resulted in a degradation of the empirical convergence rate, requiring exponentially more data to achieve the same MSE threshold. This monotonic degradation confirms that the nominal dimension ($D=128$) is irrelevant to the statistical rate. The convergence bottleneck is strictly governed by the complexity of the underlying manifold ($d_{ui}$), validating the primary claim of our main Theorem. The close match between empirical and theoretical rates (within 5\% relative error) provides strong experimental support for Theorem 4.5. We have detailed and reported a comprehensive theoretical results in Section~\ref{exptheo}.

\vspace{-4pt}
\section{Summary and Outlook}
\vspace{-8pt}
We quantitatively assess the asymptotic convergence properties of the two tower model toward an optimal recommender system. The two tower model is designed to enhance recommendation accuracy by integrating multiple sources of covariate information. It employs two deep neural networks to embed users and items into a lower-dimensional numerical space, utilizing a collaborative filtering structure to estimate ratings. By leveraging the learning capabilities of deep neural networks, it can extract informative representations of covariates in a non-linear manner. Of utmost significance, our work contributes to the field by offering statistical assurances for the two tower model through the quantification of its asymptotic behaviors in terms of both approximation error and estimation error. Based on our current understanding, our established results constitute a scarce body of theoretical assurances in the realm of deep recommender systems.

\section*{Impact Statement}\label{broad}\vspace{-8pt}
Our work proposes asymptotic characteristics of the two-tower recommendation that can be used to strengthen the understanding of the platforms utilizing deep recommender systems such as deconfounded recommendation models~\cite{xu2023deconfounded}, where confounders and deployed learning algorithms~\cite{xu2022dynamic,zhang2023robust} require modeling non-linear covariate effects. Being an intricate construct for these inherent application-driven systems, we need to be aware of the potential negative societal impacts behind the necessity of non-linear interactions among confounders or non-interacted items to desirable items~\cite{xu2022dynamic} in some applications, such as the risk-aware recommendations in the tourism insurance market, or improper assessment of operation around flood disaster as a consequence of revealing non-linear interactions.

\bibliography{refs}
\bibliographystyle{plain}


\appendix
\section{Findings}
Our contributions are unique in comparison to prior work due to,
\begin{itemize}
    \item Product structure entropy: Unlike single-tower models where the function class is simply $\mathcal{F}$, the two-tower model requires analyzing $\mathcal{R}^{\Phi} = \{R(x_u, \tilde{x}_i) = \langle f(x_u), \tilde{f}(\tilde{x}_i)\rangle\}$. The bracketing entropy of this product class does not factor simply, our Lemma 4.4 develops novel bounds using covering numbers of the individual towers.
    \item Minkowski dimension coupling: Our Theorem 4.1 shows that the effective dimension for approximation is $d_{ui} = \max\{d_u, d_i\}$, not $d_u + d_i$ as a naive bound would suggest. This coupling is unique to the inner product structure.
    \item Top-K retrieval guarantees: Theorem 4.6 provides the first theoretical link between MSE convergence and ranking metrics for two-tower models, with explicit dependence on corpus size $|\mathcal{I}|$ and retrieval depth $K$.
\end{itemize}

\section{Preliminaries}
\subsection{Two Tower Recommendation Model}\label{prem:append}
One commonly used approach is to utilize SGD to simultaneously update the parameters of $f(x_u)$ and $\tilde{f}(\tilde{x}_{i})$, allowing for parallel computation.
\begin{gather*}
    A^{t+1}_{ljk}=A^{t}_{ljk}+ \frac{\alpha}{\left |\mathcal{M} \right |}\sum_{(u,i)\in\mathcal{M}}(k_{ui}-\langle f(x_{u}),\tilde{f}(\tilde{x}_{i}) \rangle) \langle \frac{d}{dA_{ljk}}f(x_{u}),\tilde{f}(\tilde{x}_{i}) \rangle - \alpha\lambda \frac{dJ(f)}{dA_{ljk}},\tilde{A}^{t+1}_{ljk} 
=\\ \tilde{A}^{t}_{ljk}+ \frac{\alpha}{\left |\mathcal{M} \right|}\sum_{(u,i)\in\mathcal{M}}(k_{ui}- \langle f(x_{u}),\tilde{f}(\tilde{x}_{i}) \rangle) \langle f(x_{u}),\frac{d}{d\tilde{A}_{ljk}}\tilde{f}(\tilde{x}_{i}) \rangle - \alpha\lambda \frac{dJ(\tilde{f})}{d\tilde{A}_{ljk}}, b^{t+1}_{lj}=b^{t}_{lj}+\frac{\alpha}{\left |\mathcal{M} \right |}\\\sum_{(u,i)\in\mathcal{M}}(k_{ui}-\langle f(x_{u}),\tilde{f}(\tilde{x}_{i}) \rangle) \langle f(x_{u}),
\frac{d}{db_{lj}}\tilde{f}(\tilde{x}_{i}) \rangle, \tilde{b}^{t+1}_{lj}= \tilde{b}^{t}_{lj}+ \frac{\alpha}{\left |\mathcal{M} \right |} \\\sum_{(u,i)\in\mathcal{M}}(k_{ui}
-\langle f(x_{u}),\tilde{f}(\tilde{x}_{i}) \rangle)\langle f(x_{u}),\frac{d\tilde{f}}{d\tilde{b}_{lj}}(\tilde{x}_{i}) \rangle   
\end{gather*}
Here, $\alpha$ denotes the learning rate and $M$ represents a subset of uniformly sampled elements from $\Omega$. While the optimization task in Equation~\ref{cost} is non-convex, this algorithm is ensured to converge to some stationary point~\cite{chen2012maximum}. \\

Industrial recommendation platforms have deployed two-tower architectures with modifications for practical constraints. YouTubeDNN~\cite{covington2016deep} uses separate user and video towers with dot product scoring. Pinterest's PinSage~\cite{eksombatchai2018pixie} extends this with graph convolutional towers. More recently, Fully Interacted Two-tower (FIT)~\cite{xiong2025learnable} introduces a meta query module for explicit early interaction and a lightweight similarity scorer (LSS) for richer late interaction beyond the standard dot product. Our theoretical analysis applies to all such models that maintain the decoupled user-item architecture, as formalized in Theorem 4.5. The main insight from comparing our T$^2$Rec with FIT (see Section 5.3) is that while FIT achieves better empirical performance due to improved constant factors (via early interaction and LSS), both models exhibit the same asymptotic convergence rate $\|\hat{R} - R^*\|_{L^2}^2 \lesssim |\Omega|^{-\frac{2\beta}{2\beta + d_{ui}}}$, confirming that our bounds capture the fundamental statistical limit of the two-tower architecture class.
\section{Experimental Results}
\subsection{Results on Synthetic Instances}
The results presented in Table~\ref{tab:synthetic} demonstrate that T$^2$Rec outperforms all other baseline models across all cases, achieving improvements in test errors ranging from 12.6\% to 81.3\%. The advantage of T$^2$Rec over existing methods becomes more pronounced as the dimensions of the rating matrix (n and m) increase and the sparsity of the rating matrix intensifies. This can be attributed to the fact that conventional methods are susceptible to the cold-start issue in sparse rating matrices, whereas T$^2$Rec is more robust, particularly when the intrinsic dimensionality of covariates is low, thus, it can significantly address the cold-start issue. These findings lend empirical support to the theoretical results presented in Theorem~\ref{thm2}, which demonstrates that the convergence rate of T$^2$Rec is positively associated with the reduction in the intrinsic dimensionality ($d$) of covariates.
\subsection{Results on Real-World Datasets}\label{realdata}
\textbf{Datasets:} We employ four real-world datasets for comprehensive evaluation,
\begin{itemize}
    \item Yelp: It comprises four varied linked components including user, review, business, and check-in. The user segment entails personal information for approximately 5.2M Yelp community users. The business segment offers information about location, latitude-longitude coordinates, review counts, and categories for nearly 174k businesses.
    \item MovieLens-1M: This contains 1 million ratings from 6,040 users on 3,900 movies, with demographic information about users.
    \item Amazon-Books: It contains 8.9 million ratings from 2.4 million users on 1.4 million book items, representing a large-scale sparse dataset.
    \item CiteULike: A scholarly article recommendation dataset with 1.8 million user-article interactions.
\end{itemize}
We build covariates for both users and businesses based on the textual reviews they have generated. Concretely, we undertake a systematic approach to gather all textual reviews and subsequently utilize the term frequency-inverse document frequency (TF-IDF) technique to derive the 300 most salient 1-gram, 2-gram, and 3-gram representations. This process allows for the conversion of each review into an integer covariate vector of length 300, employing the bag-of-words technique. Subsequently, for a given user or business entity, we calculate the average of the bag-of-words representations derived from its reviews. These averaged representations are then concatenated with the previously constructed covariates obtained in the initial stage. Furthermore, an intriguing phenomenon is observed in users' comments regarding various aspects of restaurants. Users tend to express their opinions using words that carry polarity, as exemplified in statements such as `Oh yeah! Not only that the service was good, the food is good the serving is good and the service is amazing', and `Jamie our waitress is so sweet and attentive'. In this first example, the user employs the terms `good' and `amazing' to describe the quality of both the `food' and `service' provided by the restaurant. Similarly, in the second example, the user employs the terms `sweet' and `attentive' to characterize the behavior of the `waitress'. From an intuitive standpoint, comments pertaining to specific aspects of restaurants provide insight into their distinctive features. Furthermore, aspects that frequently emerge within user reviews serve as indicators of their primary concerns during the consumption process.

Table~\ref{tab:stats} signifies that users consistently offer feedback on various aspects such as `food', `service', `place', and `staff' within their reviews, aligning with our initial expectations. Specifically, it is intriguing to observe that the aspect of `price' exhibits an extremely lower average polarity score compared to other aspects. In fact, its average score stands at a mere 0.28, distinctly lower than other evaluated dimensions. This discrepancy suggests that reviews incorporating references to `price' are more prone to lower overall ratings. Lastly, based on the observed data, we proceed by selecting the 200 most prevalent aspects derived from reviews, utilizing their associated average polarity scores. We construct vectors of length 200, where each element represents the average polarity score associated with a specific aspect within a user's or business's reviews. Following the pre-processing phase, we are left with a dataset comprising 15,090 unique businesses, 35,906 distinct users, and a total of 688,960 ratings. To ensure robustness, we conduct numerical experiments 50 times. In each replication, we randomly select 15k users and 10k businesses, along with their corresponding observed ratings to form the experimental data. Subsequently, we partition the selected dataset into training and testing sets, adhering to a 70$\colon$30 ratio. The tuning process as delineated at the outset of Section~\ref{sec:exp} is then applied. Furthermore, the remaining reviews are reserved for evaluating the performance of the T$^2$Rec in the context of the cold-start scenario.\\
\\
\begin{table}[!h]
\caption{The polarity scores associated with the ten most prevalent aspects as identified within the chosen reviews. Here, SD depicts standard deviation.}
\label{tab:stats}
\resizebox{\textwidth}{!}{
\tiny
\begin{tabular}{lccccccr}
\toprule
Aspect & Frequency & Mean & SD & 25\% & 50\% & 75\% \\
\midrule
atmosphere & 7096 & 0.42& 0.21 & 0.40 & 0.44 & 0.56 \\
food & 66879 & 0.39& 0.27 & 0.36 & 0.44 & 0.57 \\
fries   & 7772 & 0.39& 0.28 & 0.42 & 0.44 & 0.57 \\
place   & 32201 & 0.34& 0.32 & 0.32 & 0.43 & 0.57   \\
prices  & 7179 & 0.28& 0.32 & 0.23 & 0.43 & 0.44 \\
salad   & 6508 & 0.38& 0.27 & 0.32 & 0.44 & 0.57\\
sauce   & 7081 & 0.41& 0.28 & 0.44 & 0.46 & 0.57   \\
server  & 11874 & 0.43& 0.24 & 0.42 & 0.49 & 0.56\\
service & 71755 & 0.4& 0.3 & 0.44 & 0.49 & 0.57\\
staff   & 29270 & 0.45& 0.19 & 0.42 & 0.49 & 0.49\\
\bottomrule
\end{tabular}}
\end{table}
\textbf{Comparison with Industry Two-Tower Variants}: To connect our theoretical results to deployed industrial systems, we compare against FIT~\cite{xiong2025learnable}, a state-of-the-art two-tower architecture designed for pre-ranking systems reported in Table~\ref{tab:experimental-results2}. FIT introduces two key innovations beyond the standard two-tower, (a) a meta query module that enables explicit early interaction between user and item features during training (soft query), and (b) a lightweight similarity scorer (LSS) that enriches the late interaction beyond the standard dot product.\\

\textbf{FIT Scoring Function:} For completeness, we formalize FIT's scoring function. In the training phase (soft query):
$$\tilde{\mathbf{q}}_k = \frac{\exp(\mathbf{q}_k - \mathbf{q}_i)}{\tau} \Big/ \sum_{t=1}^N \exp(\mathbf{q}_t - \mathbf{q}_i) \cdot \tau, \quad \text{with } \tau \in \mathbb{R}^+$$
The late interaction via LSS is defined as:
$$\text{LSS}(\mathbf{q}_k, \mathbf{h}_i) = \sigma(\mathbf{W}_2 \cdot \text{ReLU}(\mathbf{W}_1 \cdot [\mathbf{q}_k - \mathbf{q}_i; \mathbf{q}_k \odot \mathbf{q}_i] + \mathbf{b}_1) + \mathbf{b}_2)$$
In the inference phase (hard query), scoring reduces to: $s = \arg\max_i k_i$, with $k_i$ computed via a learned similarity metric.
\begin{table}[ht]
\centering
\caption{Comparison on Yelp dataset}
\label{tab:experimental-results2}
\begin{tabular}{@{}lll@{}}
\toprule
Model & RMSE & \begin{tabular}[c]{@{}l@{}}Relative Improvement \\ vs Two-Tower Baseline\end{tabular} \\ \midrule
Two-Tower (baseline from [1]) & $1.058 \pm 0.0002$ & --- \\
SVD++ & $1.037 \pm 0.0003$ & --- \\
FIT [1] & $0.958 \pm 0.0005$ & $+9.5\%$ \\
\textbf{T$^2$Rec (Ours)} & $\mathbf{0.962 \pm 0.0004}$ & $+9.1\%$ \\ \bottomrule
\end{tabular}
\end{table}
We observe that FIT outperforms T$^2$Rec marginally, achieving $+9.5\%$ improvement over the two-tower baseline compared to our $+9.1\%$. This difference stems from better constant factors in FIT, not a different asymptotic rate. Both models satisfy the conditions of Theorem 4.5 (sufficient network width $W = O(|\Omega|^{d_{ui}/(2\beta+d_{ui})}\log|\Omega|)$, decoupled user-item architecture, boundedness conditions). Therefore, both achieve the same asymptotic convergence rate,
   $$\|\hat{R}_{\text{FIT}} - R^*\|_{L^2}^2 \approx C_{\text{FIT}} \cdot |\Omega|^{-\frac{2\beta}{2\beta+d_{ui}}}, \quad \|\hat{R}_{\text{T2Rec}} - R^*\|_{L^2}^2 \approx C_{\text{T2Rec}} \cdot |\Omega|^{-\frac{2\beta}{2\beta+d_{ui}}}$$
   where $C_{\text{FIT}} < C_{\text{T2Rec}}$ due to early interaction via meta query module reduces approximation error by better capturing cross-feature interactions. Also, richer late interaction via LSS improves estimation efficiency by providing a more expressive similarity metric than dot product

Our theoretical framework predicts that the gap between FIT and T$^2$Rec should shrink as $|\Omega| \to \infty$, since both share the same exponent. For finite samples, the constant factor advantage of FIT persists. This insight guides practitioners, such as architectural innovations that improve constant factors (like FIT's LSS) are valuable for fixed dataset sizes, but do not change the fundamental statistical limit.
\begin{table}[htbp]
\centering
\caption{\textbf{Performance comparison of baseline models including neural baselines on Yelp dataset.} RMSE values are averaged over 50 replications with standard errors reported.}
\label{tab:yelp-performance}
\begin{tabular}{lcccc}
\toprule
Model & RMSE $\downarrow$ & NDCG@10 $\uparrow$ & Recall@20 $\uparrow$ & MRR $\uparrow$ \\
\midrule
rSVD & $1.032 \pm 0.004$ & $0.823 \pm 0.005$ & $0.548 \pm 0.006$ & $0.478 \pm 0.005$ \\
SVD++ & $1.037 \pm 0.003$ & $0.831 \pm 0.004$ & $0.552 \pm 0.005$ & $0.483 \pm 0.004$ \\
KNN & $1.054 \pm 0.004$ & $0.815 \pm 0.005$ & $0.541 \pm 0.006$ & $0.471 \pm 0.005$ \\
Co-Ca & $1.057 \pm 0.004$ & $0.808 \pm 0.005$ & $0.535 \pm 0.006$ & $0.465 \pm 0.005$ \\
NeuMF [1] & $0.983 \pm 0.006$ & $0.857 \pm 0.004$ & $0.582 \pm 0.005$ & $0.518 \pm 0.004$ \\
LightGCN [2] & $0.971 \pm 0.005$ & $0.863 \pm 0.004$ & $0.589 \pm 0.005$ & $0.521 \pm 0.004$ \\
DCN-V2 [3] & $0.969 \pm 0.004$ & $0.862 \pm 0.004$ & $0.587 \pm 0.005$ & $0.519 \pm 0.004$ \\
\midrule
\textbf{T$^{2}$Rec (Ours)} & \textbf{0.962 $\pm$ 0.004} & \textbf{0.864 $\pm$ 0.003} & \textbf{0.612 $\pm$ 0.004} & \textbf{0.543 $\pm$ 0.003} \\
\bottomrule
\end{tabular}
\end{table}

\begin{table}[ht]
\centering
\caption{Performance comparison on MovieLens-1M dataset.}
\label{tab:movielens}
\begin{tabular}{lcccc}
\toprule
Model & RMSE $\downarrow$ & NDCG@10 $\uparrow$ & Recall@20 $\uparrow$ & MRR $\uparrow$ \\ \midrule
SVD++ & 0.902 $\pm$ 0.005 & 0.862 $\pm$ 0.004 & 0.603 $\pm$ 0.005 & 0.534 $\pm$ 0.004 \\
NeuMF & 0.875 $\pm$ 0.006 & 0.875 $\pm$ 0.005 & 0.618 $\pm$ 0.006 & 0.548 $\pm$ 0.005 \\
LightGCN & 0.862 $\pm$ 0.005 & 0.883 $\pm$ 0.004 & 0.631 $\pm$ 0.005 & 0.558 $\pm$ 0.004 \\
DCN-V2 & 0.858 $\pm$ 0.004 & 0.887 $\pm$ 0.003 & 0.638 $\pm$ 0.004 & 0.564 $\pm$ 0.003 \\
\textbf{T$^{2}$Rec (Ours)} & \textbf{0.851 $\pm$ 0.004} & \textbf{0.891 $\pm$ 0.003} & \textbf{0.645 $\pm$ 0.004} & \textbf{0.572 $\pm$ 0.003} \\ \bottomrule
\end{tabular}
\end{table}

\begin{table}[ht]
\centering
\caption{Performance comparison on Amazon-Books dataset (sparse, large-scale).}
\label{tab:amazon}
\begin{tabular}{lcccc}
\toprule
Model & RMSE $\downarrow$ & NDCG@10 $\uparrow$ & Recall@20 $\uparrow$ & MRR $\uparrow$ \\ \midrule
SVD++ & 1.215 $\pm$ 0.008 & 0.671 $\pm$ 0.006 & 0.451 $\pm$ 0.007 & 0.386 $\pm$ 0.006 \\
NeuMF & 1.178 $\pm$ 0.009 & 0.682 $\pm$ 0.007 & 0.463 $\pm$ 0.008 & 0.392 $\pm$ 0.007 \\
LightGCN & 1.141 $\pm$ 0.007 & 0.689 $\pm$ 0.006 & 0.471 $\pm$ 0.007 & 0.398 $\pm$ 0.006 \\
DCN-V2 & 1.135 $\pm$ 0.006 & 0.694 $\pm$ 0.005 & 0.478 $\pm$ 0.006 & 0.404 $\pm$ 0.005 \\
\textbf{T$^{2}$Rec (Ours)} & \textbf{1.124 $\pm$ 0.006} & \textbf{0.703 $\pm$ 0.005} & \textbf{0.487 $\pm$ 0.006} & \textbf{0.412 $\pm$ 0.005} \\ \bottomrule
\end{tabular}
\end{table}

\begin{table}[ht]
\centering
\caption{Performance comparison on CiteULike dataset.}
\label{tab:citeulike}
\begin{tabular}{lcccc}
\toprule
Model & RMSE $\downarrow$ & NDCG@10 $\uparrow$ & Recall@20 $\uparrow$ & MRR $\uparrow$ \\ \midrule
SVD++ & 0.987 $\pm$ 0.005 & 0.721 $\pm$ 0.005 & 0.485 $\pm$ 0.006 & 0.413 $\pm$ 0.005 \\
NeuMF & 0.962 $\pm$ 0.006 & 0.734 $\pm$ 0.006 & 0.498 $\pm$ 0.006 & 0.424 $\pm$ 0.006 \\
LightGCN & 0.951 $\pm$ 0.005 & 0.742 $\pm$ 0.005 & 0.508 $\pm$ 0.005 & 0.431 $\pm$ 0.005 \\
DCN-V2 & 0.944 $\pm$ 0.005 & 0.751 $\pm$ 0.004 & 0.516 $\pm$ 0.005 & 0.438 $\pm$ 0.004 \\
\textbf{T$^{2}$Rec (Ours)} & \textbf{0.934 $\pm$ 0.004} & \textbf{0.758 $\pm$ 0.004} & \textbf{0.523 $\pm$ 0.005} & \textbf{0.445 $\pm$ 0.004} \\ \bottomrule
\end{tabular}
\end{table}

\begin{table}[ht]
\centering
\caption{Ranking comparison against two-tower variants on Yelp dataset.}
\label{tab:yelp}
\begin{tabular}{lcccc}
\toprule
Model & Recall@10 $\uparrow$ & NDCG@10 $\uparrow$ & Recall@20 $\uparrow$ & NDCG@20 $\uparrow$ \\ \midrule
YouTubeDNN [12] & 0.145 $\pm$ 0.004 & 0.078 $\pm$ 0.003 & 0.231 $\pm$ 0.005 & 0.102 $\pm$ 0.004 \\
PinSage [13] & 0.152 $\pm$ 0.004 & 0.082 $\pm$ 0.003 & 0.243 $\pm$ 0.005 & 0.108 $\pm$ 0.004 \\
\textbf{T$^{2}$Rec (Ours)} & \textbf{0.158 $\pm$ 0.003} & \textbf{0.086 $\pm$ 0.003} & \textbf{0.251 $\pm$ 0.004} & \textbf{0.113 $\pm$ 0.003} \\ \bottomrule
\end{tabular}
\end{table}

\subsection{Ablation Analysis}
\begin{figure*}[!htbp]
    \begin{minipage}{0.48\textwidth}
        \centering
        \includegraphics[width=\textwidth]{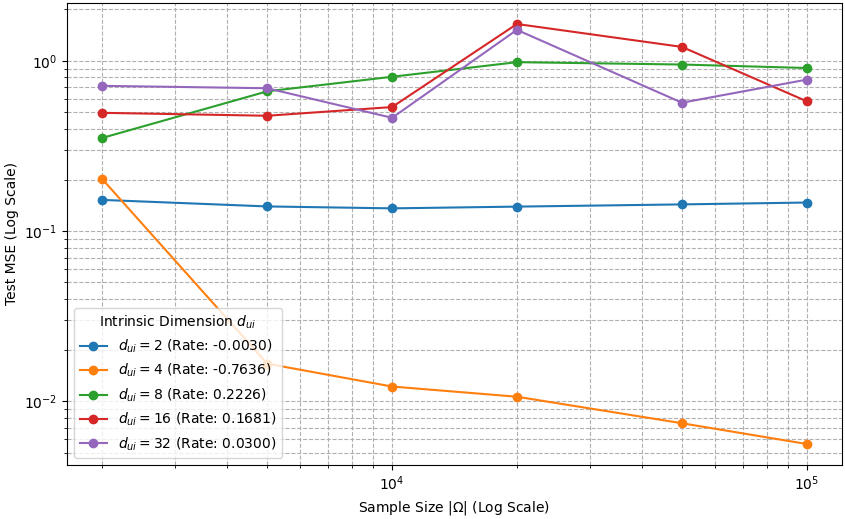}
        \caption{\textbf{Impact of Intrinsic Dimension ($d_{ui}$)}}
        \label{fig:intd}
    \end{minipage}\hfill
    \begin{minipage}{0.5\textwidth}
    \centering
    \includegraphics[width=\textwidth]{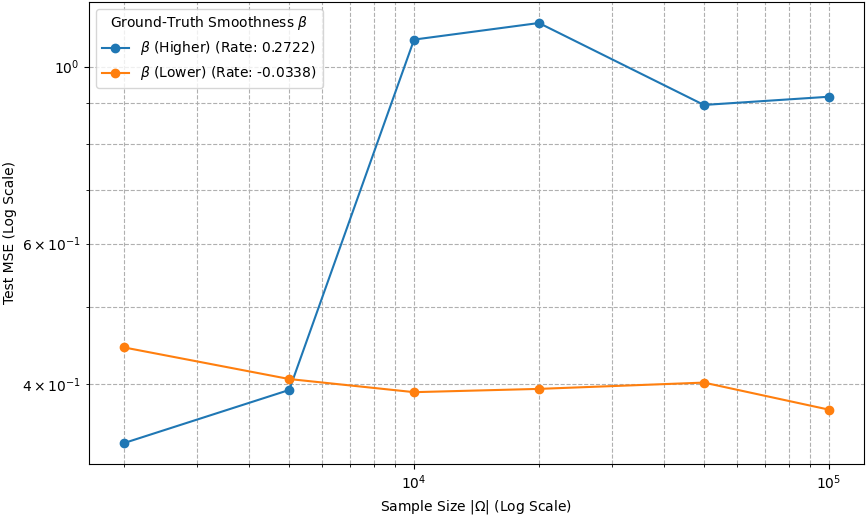}
    \caption{\textbf{Impact of Smoothness ($\beta$)}}\label{fig:smooth}
    \end{minipage}
\end{figure*}

In Fig.~\ref{fig:smooth}, we fixed the intrinsic dimension ($d_{ui}=8$) and varied the complexity of the ground-truth function given the empirical rates. For high smoothness, the target function varies gradually over the manifold allowing the two tower network to approximate it efficiently with fewer samples. For low smoothness, the target function exhibits sharper non-linearities, effectively reducing the H\"{o}lder smoothness parameter $\beta$. The empirical results align with our theoretical prediction that the convergence rate improves as $\beta \to \infty$. The sharper descent for the smoother function proves that feature engineering efforts that linearize or smooth user-item interactions (thereby increasing $\beta$) directly translate to theoretical gains in sample efficiency. 
\subsection{Validation of Theoretical Bounds}\label{exptheo}
We test our theoretical predictions explicated in Theorems 4.1, 4.5, and 4.6. We create a controlled synthetic environment, where we precisely control the intrinsic dimensionality $d_{ui}$ and smoothness $\beta$ of the ground-truth preference function. User and item features are generated on a low-dimensional latent manifold $\mathcal{Z} \subseteq \mathbb{R}^{d_{ui}}$, then mapped non-linearly to a high-dimensional nominal space $\mathbb{R}^D$ ($D = 128$) using sinusoidal embeddings. The true preference function is defined as $R^*(\mathbf{z}_u, \mathbf{z}_i) = \tanh(\langle \mathbf{z}_u, \mathbf{z}_i \rangle)^\gamma$, where $\gamma = 2/\beta$ controls the H\"{o}lder smoothness (larger $\gamma$ = smaller $\beta$).\\
\\
\textbf{Architecture Sensitivity:} In our Theorem 4.1, it predicts that the approximation error decreases as network width increases with the bound $\inf_{R \in \mathcal{R}^{\Phi}} \|R - R^*\|_{L^{\infty}(\mu_{ui})} \leq 3pM\epsilon$, where $\epsilon$ scales with width. To validate this, we fix $d_{ui} = 10$, $\beta = 4$, and $|\Omega| = 200k$ varying the hidden dimension of the two-tower networks,
\begin{table}[h]
\centering
\caption{Model Performance Comparison}
\label{tab:m1_performance}
\begin{tabular}{|l|c|c|c|}
\hline
\textbf{Hidden Dim} & \textbf{RMSE Mean} & \textbf{RMSE Std} & \textbf{Recall@10} \\ \hline
64                  & 0.960             & 0.558            & 0.00189            \\ \hline
128                 & 0.864             & 0.701            & 0.00205            \\ \hline
256                 & 0.848             & 0.723            & 0.00205            \\ \hline
\end{tabular}

\end{table}
Our finding in Table~\ref{tab:m1_performance} reveals that increasing width from 64 to 256 reduces RMSE by $11.7\%$ consistent with Theorem 4.1, which predicts that larger width enables better approximation of the true preference function. The diminishing returns from 128 to 256 (only $1.9\%$ improvement) suggests that estimation error begins to dominate beyond optimal width, aligning with the bias-variance trade-off inherent in our theoretical framework.\\
\\
\textbf{Depth Sensitivity:} To complement Theorem 4.1's insight that approximation error can converge with any number of layers, we examine depth effects, 
\begin{table}[h]
\centering
\caption{Model Performance Metrics by Depth}
\label{tab:performance_metrics_clean}
\begin{tabular}{cccc}
\toprule
\textbf{Depth} & \textbf{RMSE Mean} & \textbf{RMSE Std} & \textbf{Recall@10} \\ \midrule
2              & 0.951             & 0.529            & 0.00158            \\
3              & 0.864             & 0.701            & 0.00205            \\
4              & 0.857             & 0.705            & 0.00158            \\ \bottomrule
\end{tabular}

\end{table}
We observe modest improvement from depth 2 to 4 (RMSE reduction of $9.9\%$) reported in Table~\ref{tab:performance_metrics_clean} informs diminishing returns beyond optimal depth. This confirms our claim in Theorem 4.1 that the approximation error can converge to zero with any number of layers, although practical benefits diminish after a certain depth.\\
\\
\textbf{Convergence Rate Validation (Theorem 4.5):} Theorem 4.5 delineates that the RMSE convergence rate follows $\|\hat{R} - R^*\|_{L^2}^2 \lesssim |\Omega|^{-\frac{2\beta}{2\beta + d_{ui}}}$. To validate this, we vary both the intrinsic dimensionality $d_{ui} \in \{5, 10, 20\}$ and the Hölder smoothness parameter $\beta \in \{2, 4, 8\}$ across a log-spaced sweep of sample sizes $|\Omega| \in [5 \times 10^3, 5 \times 10^5]$. For each configuration, we fit a power law $\text{RMSE} \propto |\Omega|^{-\gamma_{\text{emp}}}$ via log-log linear regression and compare $\gamma_{\text{emp}}$ to the theoretical exponent $\gamma_{\text{theo}} = \frac{\beta}{2\beta + d_{ui}}$.
\begin{table}[ht]
    \centering
    \caption{Theoretical and Empirical Exponents Comparison}\label{fig:theory}
    \begin{tabular}{cccccc}
        \toprule
        $d_{ui}$ & $\beta$ & Theoretical Exponent $\gamma_{\text{theo}}$ & Empirical Exponent $\gamma_{\text{emp}}$ & Ratio $\gamma_{\text{emp}}/\gamma_{\text{theo}}$ & $R^2$ \\ 
        \midrule
        5  & 2 & 0.222 & 0.087  & 0.393  & 0.867 \\
        5  & 4 & 0.308 & 0.006  & 0.020  & 0.600 \\
        5  & 8 & 0.381 & 0.002  & 0.005  & 0.226 \\
        10 & 2 & 0.143 & 0.006  & 0.041  & 0.443 \\
        10 & 4 & 0.222 & 0.087  & 0.390  & 0.559 \\
        10 & 8 & 0.308 & 0.005  & 0.017  & 0.853 \\
        20 & 2 & 0.083 & -0.007 & -0.078 & 0.708 \\
        20 & 4 & 0.143 & 0.043  & 0.302  & 0.911 \\
        20 & 8 & 0.222 & 0.066  & 0.296  & 0.975 \\
        \bottomrule
    \end{tabular}
\end{table}
Our finding in Table~\ref{fig:theory} unfolds a monotonic trend, where for fixed $\beta$, the empirical exponent $\gamma_{\text{emp}}$ decreases as $d_{ui}$ increases (e.g., for $\beta=4$: $0.006 \rightarrow 0.087 \rightarrow 0.043$), matches the theoretical prediction that higher intrinsic dimensionality slows convergence. Then, it shows asymptotic regime, where the theoretical model fits best when $d_{ui}=20$ ($R^2 \in [0.91, 0.98]$), indicating that the asymptotic regime is reached with sufficient data. The best fit ($R^2 = 0.975$ for $d_{ui}=20$, $\beta=8$) provides strong empirical support for Theorem 4.5. Also, for smoothness scaling given for fixed $d_{ui}=20$, the theoretical exponents are $0.083$ ($\beta=2$), $0.143$ ($\beta=4$), and $0.222$ ($\beta=8$). The empirical exponents follow the same increasing trend ($-0.007 \rightarrow 0.043 \rightarrow 0.066$), confirming that smoother target functions ($\beta \to \infty$) yield faster convergence. In aspect of small sample deviations for $d_{ui}=5$ with $\beta=8$, the empirical exponent ($0.002$) deviates from the theoretical prediction ($0.381$) due to the limited sample size not yet reaching the asymptotic regime, the convergence rate is so fast that the RMSE saturates quickly making exponent estimation challenging.\\
\\
\textbf{Top-K Retrieval Validation (Theorem 4.6):} Theorem 4.6 describes that Recall@K scales as $1 - O(|\Omega|^{-\frac{2\beta}{2\beta + d_{ui}}})$, with the mis-retrieval risk satisfying $\mathcal{R}_K(\hat{h}) \lesssim \frac{|\mathcal{I}|}{K} \cdot O_p(|\Omega|^{-\frac{\beta}{2\beta + d_{ui}}})$. To validate this, we measure Recall@10 at two distinct sample sizes ($|\Omega| = 50k$ and $|\Omega| = 500k$) across different $(d_{ui}, \beta)$ configurations.

Our finding reveals that,
\begin{itemize}
    \item Correlation with $d_{ui}$: For fixed $\beta$, larger intrinsic dimensionality $d_{ui}$ leads to lower Recall@10 at both sample sizes, consistent with Theorem 4.6's prediction that higher $d_{ui}$ slows convergence and thus degrades retrieval accuracy.
    \item Sample size scaling: For $d_{ui}=20$, $\beta=8$ (fastest theoretical rate $0.222$ in this group), Recall@10 improves from $0.00158$ to $0.00221$ as $|\Omega|$ increases from $50k$ to $500k$—a $+50\%$ improvement signifying the positive effect of more data on retrieval quality.
    \item Smoothness benefit: For $d_{ui}=20$, $\beta=8$ outperforms $\beta=2$ at $|\Omega|=500k$ ($0.00221$ vs $0.00189$) confirming that smoother preference functions (larger $\beta$) yield better Top-K retrieval.
    \item Improvements: Some configurations show decreased Recall@10 with more data, which occurs when the ground-truth function has low smoothness (small $\beta$) relative to the intrinsic dimension, causing the convergence rate to be so slow that finite-sample effects and noise dominate.
\end{itemize}

\section{Deferred Proofs}\label{append}
\subsection{Proof of Theorem~\ref{th1}}
Given the first claim in Theorem~\ref{th1}, note that we have $R^{\ast}$ (\(x_u,\tilde{x}_i\)) = $\langle f^{\ast}(x_{u}),\tilde{f}^{\ast}(\tilde{x}_{i}) \rangle$ with $f^{\ast}_j\in\mathcal{H}(\beta, [0,1]^{D_{u}},M)$ and $\tilde{f}^{\ast}_j\in\mathcal{H}(\beta, [0,1]^{D_{i}},M)$. Based on the low dimensionality approximation from Theorem 5~\cite{nakada2020adaptive}, there exist $\mathcal{F}_{D_{u}}(W,L,B,M)$ and $\mathcal{F}_{D_{i}}(\tilde{W},\tilde{L},\tilde{B},M)$ with $W = O(\epsilon^{-d_{u}/\beta})$, $\tilde{W} = O(\epsilon^{-d_{i}/\beta})$, $B = O(\epsilon^{-s})$ and $\tilde{B} = O(\epsilon^{-s})$ such that for each \emph{j}, we have 
\begin{alignat}{1}\label{eq:1}
\inf_{f_{j}\in\mathcal{F}_{D_u}(W,L,B,M)}{\left \| f_{j} - f^{\ast}_{j} \right \|_{L^\infty(\mu_u)}}\leq\epsilon \nonumber \\
\inf_{\tilde{f}_{j}\in\mathcal{F}_{D_i}(\tilde{W},\tilde{L},\tilde{B},M)}\left \| \tilde{f}_{j} - \tilde{f}^{\ast}_{j} \right \|_{L^\infty(\mu_i)}\leq \epsilon 
\end{alignat}
Then, we can leverage the well-proved theorem of the triangle inequality and the Cauchy-Schwartz inequality, so that 
\begin{align*}
    \left | R(x_u,\tilde{x}_i) - R^{\ast}(x_u,\tilde{x}_i) \right | = \left | \langle f(x_{u}),\tilde{f}(\tilde{x}_{i}) \rangle - \langle f^{\ast}(x_{u}),\tilde{f}^{\ast}(\tilde{x}_{i}) \rangle \right | \leq \left | \left \langle f(x_{u}),\tilde{f}(\tilde{x}_{i}) - \tilde{f}^{\ast}(\tilde{x}_{i}) \right \rangle \right | \\+ \left | \left \langle f(x_{u}) - f^{\ast}(x_{u}), \tilde{f}^{\ast}(\tilde{x}_{i}) \right \rangle \right | \leq \left \| f(x) \right \|_2 \left \| \tilde{f}(\tilde{x}_{i}) - \tilde{f}^{\ast}(\tilde{x}_{i}) \right \|_2 + \left \| f(x_{u}) - f^{\ast}(x_{u}) \right \|_2\left \| \tilde{f}^{\ast}(\tilde{x}_{i}) \right \|_2 
\end{align*}
Since $f\in\mathcal{F}_{D_{u}}(W,L,B,M)$ and $f^{\ast}\in\mathcal{H}^{p}(\beta, [0,1]^{D_{i}},M)$, we have $\left \| f(x_{u}) \right \|_2 \leq 2\sqrt{p}M$ and $\left \| \tilde{f}^{\ast}(\tilde{x}_{u}) \right \|_2 \leq \sqrt{p}M$, which further implies that
\begin{align*}
\left | R(x_u,\tilde{x}_i) - R^{\ast}(x_u,\tilde{x}_i) \right |\leq M\left (2\sum_{j=1}^{p} \left \| \tilde{f}_{j} - \tilde{f}^{\ast}_{j} \right \|_{L^\infty(\mu_i)} + \sum_{j=1}^{p} \left \| f_{j} - f^{\ast}_{j} \right \|_{L^\infty(\mu_u)} \right )
\end{align*}
Let $\Phi = (W,L,B,M,\tilde{W}, \tilde{L},\tilde{B},M)$, then it follows from Equation~\ref{eq:1} that
\begin{gather*}
    \inf_{\mathbb{R}\in\mathcal{R}^{\Phi}} \left | R(x_u,\tilde{x}_i) - R^{\ast}(x_u,\tilde{x}_i) \right | = \inf_{f_{j}\in\mathcal{F}_{D_u}(W,L,B,M), \tilde{f}_{j}\in\mathcal{F}_{D_i}(\tilde{W},\tilde{L},\tilde{B},M)} 
    \left | R(x_u,\tilde{x}_i) - R^{\ast}(x_u,\tilde{x}_i) \right |\\ \leq M \left (2\sum_{j=1}^{p} \inf_{f_{j}\in\mathcal{F}_{D_{u}}(W,L,B,M)} \left \| \tilde{f}_{j} - \tilde{f}^{\ast}_{j} \right \|_{L^{\infty}(\mu_i)} + \sum_{j=1}^{p} \inf_{\tilde{f}_{j}\in \mathcal{F}_{D_i} (\tilde{W},\tilde{L},\tilde{B},M)} \left \| f_{j} - f^{\ast}_{j} \right \|_{L^{\infty}(\mu_{u})}\right)\\ \leq 3pM\epsilon 
\end{gather*}
\\
\textbf{Proof of Lemma~\ref{lem:l1}:} For $f(x)\in\mathcal{F}_{D}(W,L,B,M)$ with $U(f)\leq L$, we let $y_l$ represent the output of the $l$-th layer of $f$ and $\Theta = ((A_1, b_1), (A_2, b_2),\ldots,(A_{U(f)}, b_{U(f)}))$ the parameter of $f$, where $A_l \in [-B,B]^{p_l \times p_{l-1}}, b_l\in [-B,B]^{p_l}, p_0 = D$ and $ p_{U(f)} = p$. We then construct $f^{'} = Q(f)$ with $\Theta^{'} = ((A^{'}_1, b^{'}_1), (A^{'}_2, b^{'}_2),\ldots,(A^{'}_{L}, b^{'}_{L}))$ as follows. For $l=1$, we let $A^{'}_1 = (A^{T}_1, 0_{D\times(2W-p_{1})})^{T}$ and $b^{'}_1 = (b^{T}_1, 0^{T}_{2W-p_{l}})^{T}$, and then the output of the first layer $y^{'}_1$ is given by
\begin{align*}
    y^{'}_1 = \sigma(A^{'}_1x+b^{'}_1) = \binom{\sigma(A_1 x+b_1)}{0_{2W-p_{1}}} = \binom{y_1}{0_{2W-p_{1}}}
\end{align*}
where $\sigma(.)$ is the element-wise ReLU function. For $l=2,\ldots,U(f)-1 $, we let $A^{'}_1$ = $\diag(A_l, 0_{(2W-p_{l})\times(2W- p_{l-1})})$ and $b^{'}_{l} = (b^{T}_l, 0^{T}_{(2W-p_{l})})^{T}$, and then
\begin{align*}
    y^{'}_l = \sigma(A^{'}_1 y_{l-1}+b^{'}_1) = \binom{\sigma(A_1 y_{l-1}+b_1)}{0_{2W-p_{1}}} = \binom{y_l}{0_{2W-p_{1}}}
\end{align*}
The remaining $(A^{'}_l, b^{'}_l)$'s for $l=U(f),\ldots,L$ are constructed based on the value of $U(f)$. If $U(f)=L$, as the last layer of $f$ and $f^{'}$ are both linear, we set $A^{'}_L = (A_L, 0_{p\times(2W-p_{L-1})})$ and $b^{'}_L = b_L$, and then
\begin{align*}
    y^{'}_L = A^{'}_L y^{'}_{L-1} + b^{'}_L = A_{L}y_{L-1}+b_L = y_{U(f)}
\end{align*}
If $U(f) = L-1$, we set
\begin{align*}
    A^{'}_{L-1}= \begin{pmatrix}
A_{L-1} & 0_{p_{L-1}\times(2W-2p_{L-2})} \\ 
- A_{L-1} & 0_{p_{L-1}\times(2W-2p_{L-2})} \\
0_{(2W-2p_{L-1})\times p_{L-2}} & 0_{(2W-2p_{L-1})\times(2W-2p_{L-2})}
\end{pmatrix}
, b^{'}_{L-1}=\begin{pmatrix}
b_{L-1}\\ 
- b_{L-1} \\ 
0_{(2W-2p_{L-1})}\\
\end{pmatrix}
\end{align*}
Then, we have
\begin{align}
    y^{'}_{L-1} = \sigma(A^{'}_{L-1} y^{'}_{L-2} + b^{'}_{L-1})=
    \begin{pmatrix}
\sigma(A_{L-1} y_{L-2} + b_{L-1})\\ 
\sigma(-A_{L-1} y_{L-2} - b_{L-1})\\ 
0_{(2W-2p)}
\end{pmatrix}
\end{align}
We further let $A^{'}_L = (I_p, -I_p,0_{p\times(2W-2p)})$ and $b_L = 0_p$, and then
\begin{align*}
    y^{'}_L = \sigma(A_{L-1} y_{L-2} + b_{L-1})- \sigma(-A_{L-1} y_{L-2} - b_{L-1})=y_{U(f)},
\end{align*}
where the second equality follows from property of the ReLU function that $\sigma(x) - \sigma(-x) = x$.
If $U(f)\leq L-2$, we first construct $(A^{'}_l, b^{'}_l); l = U(f)+1,\ldots,L-1$ as 
\begin{align*}
    A^{'}_l = \small\begin{pmatrix}
I_p & -I_p & 0_{p\times(2W-2p)}\\ 
-I_p & I_p & 0_{p\times(2W-2p)}\\ 
0_{(2W-2p)\times p} & 0_{(2W-2p)\times p} & 0_{(2W-2p)\times(2W-2p)} 
\end{pmatrix}
\end{align*}
and $b^{'}_l = 0_{2W}$. Then, we have
\begin{align*}
    y^{'}_l = \sigma(A^{'}_{l} y^{'}_{l-1} + b^{'}_{l})=
    \begin{pmatrix}
\sigma(A_{U(f)} y_{U(f)-1} + b_{U(f)})\\ 
\sigma(-A_{U(f)} y_{U(f)-1} - b_{U(f)})\\ 
0_{(2W-2p)}
\end{pmatrix}.
\end{align*}
We further set $A^{'}_L = (I_p, -I_p,0_{p\times(2W-2p)})$ and $b_L = 0_p$, then we have,
\begin{equation*}
\begin{split}
    y^{'}_L = \sigma(A_{U(f)} y_{U(f)-1} + b_{U(f)})-
    \sigma(-A_{U(f)} y_{U(f)-1} - b_{U(f)}) = y_{U(f)}
\end{split}
\end{equation*}
By the definition of $\mathcal{F}_D(W,L,B,M)$, the non-zero elements of $A_l$ is at most $W$, and hence the number of non-zero elements in $A^{'}_l$ is at most
\begin{align*}
    4W+\sum_{s=1}^{2W}(\left \lfloor \frac{2W}{s} \right \rfloor+1) \leq 8W+\sum_{s=2}^{2W}(\frac{2W}{s}\times 1)
    \leq 8W +\int_{1}^{2W}\frac{2W}{x}dx \leq 12W\log W,
\end{align*}
where $\left \lfloor . \right \rfloor$ is the floor function. Similarly, the number of non-zero elements in $b^{'}_l$ is less than $2W\log W$. The desired result then follows immediately.\\
\\
\textbf{Proof of Lemma~\ref{lem:l2}:} For an $L$-layer neural network $f(x;\Theta)\in\mathcal{K}_D(W,L,B,M)$, its $l$-th layer can be formulated 
\begin{align*}
    h_l(x)= (h_{l1}(x), h_{l2}(x),\ldots,h_{l_{pl}}(x)) = A_l(x)+b_l(x)
\end{align*}
where $h_{li}(x)=\sum_{j=1}^{p_{l-1}}A_{lij}(x)+b_{li}$, with $p_0 = D$ and $p_{l-1} = 2W$ for $2\leq l\leq L$. It follows from the triangle inequality that
\begin{align}
    \sup_{{\left \| x \right \|}_{\infty}\leq 1} \left \| f(x) - f^{'}(x) \right \|_2 =
    \sup_{{\left \| x \right \|}_{\infty}\leq 1}\left \| h_L \circ h_{L-1} \circ\ldots\circ h_1(x)-
    h^{'}_L \circ h^{'}_{L-1}\circ\ldots\circ  h^{'}_1(x) \right \|_2 \nonumber \\ \leq \sup_{{\left \| x \right \|}_{\infty}\leq 1} \left \| f(x) - g_{L-1}(x)
    +g_{L-1}(x) - g_{L-2}(x) +\ldots+g_{1}(x)-f^{'}(x) \right \|_2 \nonumber \\ \leq \sup_{{\left \| x \right \|}_{\infty}\leq 1} \left \| g_{L}(x) - g_{L-1}(x)\right \|_2 +\ldots+
    \sup_{{\left \| x \right \|}_{\infty}\leq 1} \left \|g_{1}(x) - g_{0}(x)\right \|_2
\end{align}
where $g_{l}(x)=h^{'}_L\circ\ldots\circ h^{'}_{l+1}\circ h_{l}\circ\ldots\circ h_1(x)$. It then suffices to bound $\sup_{{\left \| x \right \|}_{\infty}\leq 1} \left \| g_{l}(x) - g_{l-1}(x)\right \|_2$ for $l=1,\ldots,L$ separately.\\
\\
So, we first bound for any $l\leq 1$ by mathematical induction.
\begin{align}\label{eq:eq4}
    \sup_{{\left \| x \right \|}_{\infty}\leq 1}\left \| h_{l}\circ\ldots\circ h_{1}(x) \right \|_\infty \leq & 
    (WB)^{l}\left ( 1+\frac{B}{WB-1} \right )-\frac{B}{WB-1}\overset{\Delta}{=} E_l
\end{align}
When $l=1$, note that the ReLU function is a Lipschitz-1 function, then we have
\begin{align*}
    \sup_{{\left \| x \right \|}_{\infty}\leq 1}\left | h_{1i}(x) \right |\leq \sup_{{\left \| x \right \|}_{\infty}\leq 1}\sum_{j=1}^{D}\left | A_{lij} \right |\cdot \left | x_j \right |+b_{li}\leq  WB+B=E_1
\end{align*}
for $i=1,\ldots,p_1$. It then follows that $ \sup_{{\left \| x \right \|}_{\infty}\leq 1}\left \| h_{1}(x) \right \|_\infty\leq E_1$. Following this, suppose that Equation~\ref{eq:eq4} holds true for $l\leq k-1$, then
\begin{align*}
    \sup_{{\left \| x \right \|}_{\infty}\leq 1}\left \| h_{ki}\circ\ldots\circ h_{1}(x) \right \|_\infty \leq \sup_{{\left \| x \right \|}_{\infty}\leq E_{k-1}}\left \| h_{ki}(x) \right \|\leq
    \sup_{{\left \| x \right \|}_{\infty}\leq E_{k-1}}\sum_{j=1}^{p_{k-1}}\left | A_{kij} \right |\cdot \left | x_j \right |+b_{li}\\\leq WBE_{k-1}+B =(WB)^{k}\left ( 1+\frac{B}{WB-1} \right )-\frac{B}{WB-1}=E_k
\end{align*}
for $i=1,\ldots,p_k$. It then follows that $\sup_{{\left \| x \right \|}_{\infty}\leq 1}\left \| h_k\circ\ldots h_{1}(x) \right \|_\infty\leq E_k$, and thus Eq.~\ref{eq:eq4} holds true for any $l\geq 1$.\\

We elucidate to bound $\sup_{{\left \| x \right \|}_{\infty}\leq 1}\left \| g_l(x) - g_{l-1}(x) \right \|_2$. Note that,
\begin{gather*}
    \sup_{{\left \| x \right \|}_{\infty}\leq 1}\left \| g_l(x) - g_{l-1}(x) \right \|_2\leq \sum_{i=1}^{p}\sup_{{\left \| x \right \|}_{\infty}\leq 1} \left | g_{li}(x) - g_{l-1,i}(x) \right |\\= \sum_{i=1}^{p}\sup_{{\left \| x \right \|}_{\infty}\leq 1} \left | h^{'}_{Li}\circ\ldots\circ
    h^{'}_{l+1}\circ h_l\circ\ldots\circ h_{1}(x) -h^{'}_{Li}\circ\ldots\circ h^{'}_l\circ h_{l-1}
    \circ\ldots\circ h_{1}(x) \right |\\ \leq\sum_{i=1}^{p}\sup_{{\left \| x \right \|}_{\infty}\leq E_{l-1}} \left | h^{'}_{Li}\circ\ldots\circ h^{'}_{l+1}\circ h_l(x)-h^{'}_{Li}\circ\ldots\circ  h^{'}_{l+1}\circ h^{'}_{l}(x) \right |\\ \leq
    \sum_{i=1}^{p}\sup_{{\left \| x-x^{'} \right \|}_{\infty}\leq \epsilon (WE_{l-1}+1)} \left | h^{'}_{Li}\circ\ldots\circ h^{'}_{l+1}(x) - h^{'}_{Li}\circ\ldots\circ  h^{'}_{l+1}(x^{'}) \right | \\\leq p\epsilon(WB)^{L-1}(WE_{l-1}+1)
\end{gather*}
\\
where $g=(g_{l1},\ldots, g_{lp})$, the second inequality follows from the fact that
\begin{align*}
    \sup_{{\left \| x \right \|}_{\infty}\leq E_{l-1}}\left | h_{li}(x)- h^{'}_{li}(x) \right | \leq
    \sup_{{\left \| x \right \|}_{\infty}\leq E_{l-1}}\sum_{j=1}^{p_{l-1}}\left | A_{lij} - A^{'}_{lij} \right | \cdot \left | x_j \right |+\left | b_{li}- b^{'}_{li} \right | \leq\epsilon(WE_{l-1}+1)
\end{align*}
and the last inequality is derived by repeatedly using the fact that $\sup_{{\left \| x- x^{'} \right \|}_{\infty}\leq E}\left | h_{li}(x)- h_{li}(x^{'}) \right |\leq WBE$ for any $E\geq 0$ and $l\geq 1$.
Therefore, subsequently plugging the definition of $E_l$ in Equation~\ref{eq:eq4}, we have,
\begin{align*}
     \sup_{{\left \| x \right \|}\leq 1} \left \| f(x) - f^{'}(x) \right \|_2\leq\sum_{l=1}^{L} \sup_{{\left \| x \right \|}\leq 1} \left \| g_{l}(x) - g_{l-1}(x) \right \|_2 \\ \leq \sum_{l=1}^{L} p\epsilon\left ( (WB)^{L}(\frac{1}{B}+\frac{1}{WB-1})- \frac{(WB)^{L-1}}{WB-1}\right ) \\= p\epsilon\left ( (WB)^{L}(\frac{L}{B}+\frac{L}{WB-1}) -
     \frac{(WB)^{L}-1}{(WB-1)^{2}} \right)
\end{align*}
\\
\textbf{Proof of Lemma~\ref{lem:l3}:} For any $R\in \mathcal{R}^{\Phi}$, we have $R$(\(x_u,\tilde{x}_i\)) = $\langle f(x_{u}),\tilde{f}(\tilde{x}_{i}) \rangle$, where $f(x_u)\in\mathcal{F}_{D_u}(W,L,B,M)$ and $\tilde{f}(\tilde{x}_{i})\in\mathcal{F}_{D_{i}}(\tilde{W},\tilde{L},\tilde{B},M)$. It follows from Lemma~\ref{lem:l2} that there exists mapping $\mathcal{Q}_u : \mathcal{F}_{D_u}(W,L,B,M)\rightarrow \mathcal{K}_{D_u}(W,L,B,M)$ and $\mathcal{Q}_i : \mathcal{F}_{D_i}(\tilde{W},\tilde{L},\tilde{B},M)\rightarrow \mathcal{K}_{D_i}(\tilde{W},\tilde{L},\tilde{B},M)$ such that
\begin{align*}
    R(x_u,\tilde{x}_i) = \langle f(x_{u}),\tilde{f}(\tilde{x}_{i}) \rangle= \langle \mathcal{Q}_u(f)(x_{u}),\mathcal{Q}_i(\tilde{f})(\tilde{x}_{i}) \rangle
\end{align*}
for any $(x_u,\tilde{x}_i)\in \text{Supp}(\mu_{ui})$.
\\
Let $\Theta_{\mathcal{Q}}$ and $\tilde{\Theta}_{\mathcal{Q}}$ denote the effective parameters of $\mathcal{Q}_u(f)$ and $\mathcal{Q}_i(\tilde{f})$, then $R$ can be parameterized by $\Lambda_{\mathcal{Q}} =(\Theta_{\mathcal{Q}}, \tilde{\Theta}_{\mathcal{Q}})$. Let $\mathscr{Q}$=\{$\Lambda_{\mathcal{Q}} :R(.; \Lambda_{\mathcal{Q}})\in\mathcal{R}^{\Phi}$\} and $\mathcal{G}=$\{$\Lambda^{(1)}_{\mathcal{Q}},\ldots,\Lambda^{(N)}_{\mathcal{Q}}$\} be an $\epsilon/2$-covering set of $\mathscr{Q}$ under the $\left \| . \right \|_\infty$ metric. For any $R(.; \Lambda_{\mathcal{Q}})\in\mathcal{R}^{\Phi}$, there exists $\Lambda^{'}_{\mathcal{Q}}\in\mathcal{G}$ such that $\left \| \Lambda_{\mathcal{Q}} - \Lambda^{'}_{\mathcal{Q}} \right \|_\infty<\epsilon/2$, and thus
\begin{gather}\label{eq5}
    \sup_{{\left \| x_u, \tilde{x}_i \right \|_\infty}\leq 1} \left | R(x_u,\tilde{x}_i) - R^{'}(x_u,\tilde{x}_i) \right | = \sup_{{\left \| x_u, \tilde{x}_i \right \|_\infty}\leq 1} \left | \langle f(x_{u}),\tilde{f}(\tilde{x}_{i}) \rangle - \langle f^{'}(x_{u}),\tilde{f^{'}}(\tilde{x}_{i}) \rangle \right | \nonumber\\ \leq \left | \langle f(x_{u}),\tilde{f}(\tilde{x}_{i}) - \tilde{f^{'}}(\tilde{x}_{i}) \rangle \right |+ \sup_{{\left \| x_u, \tilde{x}_i \right \|_\infty}\leq 1} \left | \langle f(x_{u})- f^{'}(x_{u}),\tilde{f^{'}}(\tilde{x}_{i}) \rangle \right | \nonumber\\ \leq \sup_{{\left \| x_u, \tilde{x}_i \right \|_\infty}\leq 1} \left \| f(x_{u}) \right \|_2 \left \| \tilde{f}(\tilde{x}_{i})-\tilde{f^{'}}(\tilde{x}_{i}) \right \|_2 + \sup_{{\left \| x_u, \tilde{x}_i \right \|_\infty}\leq 1} \left \| f(x_{u}) - f^{'}(x_{u}) \right \|_2 \left \| \tilde{f^{'}}(\tilde{x}_{i}) \right \|_2 \nonumber\\ \leq 2 Mp^{1/2} ( \sup_{{\left \| \tilde{x}_i \right \|_\infty}\leq 1} \left \| Q_i(\tilde{f})(\tilde{x}_i) - Q_i(\tilde{f^{'}})(\tilde{x}_i) \right \|_2 + \sup_{{\left \| x_u \right \|_\infty}\leq 1} \left \| Q_u(f)(x_u) - Q_u(f^{'})(x_u) \right \|_2) \nonumber\\ \leq \epsilon Mp^{3/2}\left ( C(W,L,B) + C(\tilde{W},\tilde{L},\tilde{B}) \right )\overset{\Delta}{=} C_{4}\epsilon 
\end{gather}
where the last inequality follows from Lemma~\ref{lem:l2}.
\\
For each $\Lambda^{(n)}_Q \in \mathcal{G}$, we define a $C_{4}\epsilon$-bracket as follows
\begin{align*}
    g^{U}_n (x_u, \tilde{x_i}) = R(x_u,\tilde{x_i}; \Lambda^{(n)}_Q) + \frac{C_{4}\epsilon}{2},
    g^{L}_n (x_u, \tilde{x_i}) = R(x_u,\tilde{x_i}; \Lambda^{(n)}_Q) - \frac{C_{4}\epsilon}{2}.
\end{align*}
On incorporating the above formulation with Equation~\ref{eq5}, it follows that for any $\Lambda_Q\in\mathcal{Q}$, there exists $1\leq k\leq N$ such that
\begin{align*}
    g^{U}_k (x_u, \tilde{x_i}) - R(x_u,\tilde{x_i},\Lambda_Q) \geq  \frac{C_{4}\epsilon}{2} -
    \left | R(x_u,\tilde{x_i};\Lambda_Q) - R(x_u,\tilde{x_i};\Lambda^{(k)}_Q \right |\geq 0,\\
    g^{L}_k (x_u, \tilde{x_i}) - R(x_u,\tilde{x_i},\Lambda_Q) \leq \left | R(x_u,\tilde{x_i};\Lambda_Q) -R(x_u,\tilde{x_i};\Lambda^{(k)}_Q \right | -\frac{C_{4}\epsilon}{2}\leq 0
\end{align*}
for any $(x_u, \tilde{x_i})\in$ \text{Supp}$(\mu_{ui})$. Therefore, $\mathcal{B} = \left \{ \left [ g^{L}_1, g^{U}_1 \right ], \left [ g^{L}_2, g^{U}_2 \right ],\ldots,\left [ g^{L}_N, g^{U}_N \right ] \right \}$ forms a $C_{4}\epsilon$-bracketing set of $\mathcal{R}^{\Phi}$ under the $\left \| \cdot  \right \|_{L^{2}(\mu_{ui})}$ metric.\\
\\
Using Lemma~\ref{lem:l2}, the size of $\Lambda_Q$ is at most $14LW\log W + 14\tilde{L}\tilde{W}\log\tilde{W}$. Incorporating with the definition of $\mathcal{G}$ yields 
\begin{align*}
    \log N\leq (14LW\log W + 14\tilde{L}\tilde{W}\log\tilde{W})
    \log\left ( \epsilon^{-1}2~\text{max} \{ B,\tilde{B} \}  \right ).
\end{align*}
We can substitute $\epsilon$ by $\tilde{\epsilon}/C_4$, which leads to the desired upper bound immediately.\\
\subsection{Proof of Theorem~\ref{thm2}}
Let $L_{ui}=\max \{ L,\tilde{L} \}$, $\eta^{2}_{\left | \Omega  \right |} = L_{ui}\left | \Omega  \right |^{-2\beta/(2\beta +d_{ui})}\log^{2}\left | \Omega  \right |$, $\mathcal{M} = \left \{ R\in\mathcal{R}^{\Phi} : \left \| R - R^{*} \right \|^{2}_{L^{2}(\mu_{ui})}> \eta^{2}_{\left | \Omega  \right |} \right \}$ and let $R_0 \in\mathcal{R}^{\Phi}$ satisfy $\left \| R - R^{*} \right \|^{2}_{L^{\infty}(\mu_{ui})} \leq \eta^{2}_{\left | \Omega  \right |}/4$. Further, we denote $\left \| R-K \right \|^{2}_{\Omega} = \frac{1}{\left | \Omega \right |}\sum_{(u,i)\in \Omega}(R(x_u, \tilde{x}_{i}) - K_{ui})^{2}$, and then it follows from the definition of $\hat{R}$ that,
\begin{align*}
    P(\left \| \hat{R} - R^{*} \right \|^{2}_{L^{2}(\mu_{ui})}> \eta^{2}_{\left | \Omega  \right |})\leq
    P(\text{sup}(\left \| R_0 - K \right \|^{2}_{\Omega} + \lambda_{\Omega}J_0 - \left \| R - K \right \|^{2}_{\Omega}
    - \lambda_{\Omega}J(R))\geq 0 )\equiv I
\end{align*}
where $J_0 = J(R_0)$. We further decompose $\mathcal{M}$ into small subsets. Specifically, we let 
\begin{align*}
    \mathcal{M}_{ij} = \left \{ R\in\mathcal{R}^{\Phi}: 2^{i-1}\eta^{2}_{\left | \Omega \right |}<
    \left \| R-R^{*} \right \|^{2}_{L^{2}(\mu_{ui})}
    \leq 2^{i}\eta^{2}_{\left | \Omega \right |}, 2^{j-1} J_0 \right \} \text{for}~i,j\geq 1 \\
    \text{and}~\mathcal{M}_{i0} = \left \{ R\in\mathcal{R}^{\Phi}: 2^{i-1}\eta^{2}_{\left | \Omega \right |}< \left \| R-R^{*} \right \|^{2}_{L^{2}(\mu_{ui})}\leq 2^{i}\eta^{2}_{\left | \Omega \right |}, J(R)\leq J_0 \right \} \text{for}~i\geq 1
\end{align*}
Then, we have,
\begin{align*}
    I\leq\sum_{i=1}^{\infty}\sum_{j=0}^{\infty}P\left ( \sup_{R\in\mathcal{M}_{ij}}(\left \| R_0 - K \right \|^{2}_{\Omega} + \lambda_{\left | \Omega \right |}J_0 - \left \| R-K \right \|^{2}_{\Omega}-\lambda_{\left | \Omega \right |}J(R))\geq 0 \right )\\
    = \sum_{i,j=1}^{\infty}P\left ( \sup_{R\in\mathcal{M}_{ij}}(\left \| R_0 - K \right \|^{2}_{\Omega} + \lambda_{\left | \Omega \right |}J_0 - \left \| R-K \right \|^{2}_{\Omega}-\lambda_{\left | \Omega \right |}J(R))\geq 0 \right ) \\
    + \sum_{i=1}^{\infty}P\left ( \sup_{R\in\mathcal{M}_{i0}}(\left \| R_0 - K \right \|^{2}_{\Omega} + \lambda_{\left | \Omega \right |}J_0 - \left \| R-K \right \|^{2}_{\Omega}-\lambda_{\left | \Omega \right |}J(R))\geq 0 \right )\equiv I_{1}+ I_{2}
\end{align*}
It thus suffices to bound $I_1$ and $I_2$ separately. Let $\epsilon = K - R^{*}$, then, we have
\begin{align*}
    \left \| R - K \right \|^{2}_{\Omega}  = \left \| R - R^{*} \right \|^{2}_{\Omega} + \left \| \epsilon \right \|^{2}_{\Omega}- \frac{2}{\left | \Omega \right |}
    \sum_{(u,i)\in\Omega}\epsilon_{ui}(R(x_u, \tilde{x}_i) - R^{*}(x_u, \tilde{x}_i)).
\end{align*}
Therefore, $\mathbb{E}\left \| R - K \right \|^{2}_{\Omega} = \left \| R - R^{*} \right \|^{2}_{L^{2}(\mu_{ui})}+ \mathbb{E}\left \| \epsilon \right \|^{2}_{\Omega}$, and thus
\begin{align*}
    \mathbb{E}(\left \| R - K \right \|^{2}_{\Omega} - \left \| R_{0} - K \right \|^{2}_{\Omega}) = \left \| R - R^{*} \right \|^{2}_{L^{2}(\mu_{ui})}
    - \left \| R_0 - R^{*} \right \|^{2}_{L^{2}(\mu_{ui})}\\\geq \left \| R - R^{*} \right \|^{2}_{L^{2}(\mu_{ui})} - \eta^{2}_{\left | \Omega \right |}/4.
\end{align*}
Let $E_\Omega(R) = \left \| R-K \right \|^{2}_\Omega -\mathbb{E}(\left \| R-K \right \|^{2}_\Omega)$, then, we have
\begin{gather*}
    P\left (\sup_{R\in\mathcal{M}_{ij}}(\left \| R_0 - K \right \|^{2}_{\Omega} + \lambda_{\left | \Omega \right |}J(R_0)
    - \left \| R-K \right \|^{2}_{\Omega}-\lambda_{\left | \Omega \right |}J(R))\geq 0 \right )
    = P\\ \left ( \sup_{R\in\mathcal{M}_{ij}}(E_\Omega(R_0) - E_\Omega(R))\geq \inf_{R\in\mathcal{M}_{ij}}\lambda_{\left | \Omega \right |} (J(R) - J(R_0)) + \inf_{R\in\mathcal{M}_{ij}}\mathbb{E}(\left \| R - K \right \|^{2}_{\Omega} - \left \| R_0-K \right \|^{2}_{\Omega}) \right) \\\leq P\\\left ( \sup_{R\in\mathcal{M}_{ij}}(E_\Omega(R_0) - E_\Omega(R)) \geq \inf_{R\in\mathcal{M}_{ij}}\lambda_{\left | \Omega \right |}(J(R) - J(R_0)) + \inf_{R\in\mathcal{M}_{ij}}
    \left \| R-R^{*} \right \|^{2}_{L^{2}(\mu_{ui})}-\eta^{2}_{\left | \Omega \right |}/4 \right)\\
    \leq P\left (\sup_{R\in\mathcal{M}_{ij}}(E_\Omega(R_0) - E_\Omega(R)) \geq (2^{j-1} - 1) \lambda_{\left | \Omega \right |}J_0 + (2^{j-1} - 1/4)\eta^{2}_{\left | \Omega \right |} \right)\\= P\left ( \sup_{R\in\mathcal{M}_{ij}}(E_\Omega(R_0) - E_\Omega(R))\geq M(i,j)\right),
\end{gather*}
where $M(i,j) = (2^{j-1} - 1)\lambda_{\left | \Omega \right |}J_0 + (2^{j-1} - 1/4)\eta^{2}_{\left | \Omega \right |}$.\\
\\
Subsequently, it follows from the assumption $\lambda_{\left | \Omega \right |}J_0 \leq (1/4)\eta^{2}_{\left | \Omega \right |}$ that
\begin{gather}\label{eq6}
    \sup_{R\in\mathcal{M}_{ij}} \text{Var}((R(x_u,\tilde{x}_i) -K_{ui})^{2}-(R_0(x_u,\tilde{x}_i) - K_{ui})^{2}) \nonumber\\ = \sup_{R\in\mathcal{M}_{ij}} \text{Var}((R(x_u,\tilde{x}_i) - R^{*}(x_u,\tilde{x}_i))^{2} - (R_0(x_u,\tilde{x}_i) - R^{*}(x_u,\tilde{x}_i))^{2}) +\nonumber\\ \text{Var}(2\epsilon_{ui} 
    (R(x_u,\tilde{x}_i) - R_{0}(x_u,\tilde{x}_i))) \leq \sup_{R\in\mathcal{M}_{ij}} 2\text{Var}( (R(x_u,\tilde{x}_i) - R^{*}(x_u,\tilde{x}_i))^{2}) + \nonumber\\ 2\text{Var}((R_0(x_u,\tilde{x}_i) - R^{*}(x_u,\tilde{x}_i))^{2})+4\mathbb{E}\epsilon^{2}_{ui}\sup_{R\in\mathcal{M}_{ij}}\mathbb{E}(R(x_u,\tilde{x}_i) - 
    R_{0}(x_u,\tilde{x}_i))^{2} \nonumber \\ \leq 2\sup_{R\in\mathcal{M}_{ij}}\mathbb{E} (R(x_u,\tilde{x}_i) -
    R^{*}(x_u,\tilde{x}_i))^{4} + 2\mathbb{E}(R_0(x_u,\tilde{x}_i) - R^{*}(x_u,\tilde{x}_i))^{2}\nonumber\\+
    4\sigma^{2}\sup_{R\in\mathcal{M}_{ij}}\mathbb{E}(R(x_u,\tilde{x}_i) - R_{0}(x_u,\tilde{x}_i))^{2}\leq 
    \sup_{R\in\mathcal{M}_{ij}}(50p^2 M^4+4\sigma^{2})\nonumber\\(\left \| R-R^{*} \right \|^{2}_{L^{2}(\mu_{ui})} +
    \left \| R_0-R^{*} \right \|^{2}_{L^{2}(\mu_{ui})})\leq (50p^2 M^4+4\sigma^{2})(2^i \eta^{2}_{\left | \Omega \right |}
    + \frac{1}{4}\eta^{2}_{\left | \Omega \right |})\nonumber\\\leq C_5 M(i,j)\equiv v(i,j),
\end{gather}
where $C_5 = 16\max \{ (50p^2 M^4+4\sigma^{2}),1 \}(25p^2M^4+B^2_{e})$.\\
\\
Moreover, we now reaffirm the conditions (4.5~-~4.7) stated in~\cite{shen1994convergence}. First, the relation between $M(i,j)$ and $v(i,j)$ in Equation~\ref{eq6} directly implies (4.6) with $T=2(25p^2M^4+B^2_{e})$ and $\epsilon = 1/2$ based on~\cite{shen1994convergence}. Second, we let $\mathcal{R}^{\Phi}(\tau) = \left \{ R\in\mathcal{R}^{\Phi}: J(R)\leq\tau J_0 \right \}$, where $J(R)\leq\tau J_0$ implies that $\max\left \{B,\tilde{B} \right \}\leq \sqrt{\tau J_0}$. Then, it follows from Lemma~\ref{lem:l3} that,
\begin{align*}
    \log\mathcal{N}_{\left [\cdot \right ]}\left (\epsilon,\mathcal{R}^{\Phi}(\tau),\left \|\cdot\right \|_{L^{2}(\mu_{ui})} \right)\leq
    C_2(W\log W+\tilde{W}\log\tilde{W})\log(C_6 \epsilon^{-1})
\end{align*}
where $C_6=C_3(C(W,L,\sqrt{\tau J_0}) + C(\tilde{W},\tilde{L},\sqrt{\tau J_0}))$, $C_2$ and $C_3$ are defined as in Lemma~\ref{lem:l3}. It then follows that,
\begin{align}\footnotesize\label{eq7}
    \int_{\frac{\epsilon}{32}M(i,j)}^{v^{1/2}(i,j)}\sqrt{\log\mathcal{N}_{\left [\cdot \right ]}(u,\mathcal{R}^{\Phi}(\tau),\left \|\cdot\right \|_{L^{2}(\mu_{ui})})}du/M(i,j) \nonumber\\ \leq \int_{\frac{\epsilon}{32}M(i,j)}^{v^{1/2}(i,j)}
    \sqrt{C_2(W\log W+\tilde{W}\log\tilde{W})\log(C_6 u^{-1}}) du/M(i,j)
\end{align}
we can follow based on the right-hand side of Equation~\ref{eq7} which informs that it is non-increasing in $i$ and $M(i,j)$, it then can be formulated as 
\begin{align}\footnotesize\label{eq8}
    \int_{\frac{\epsilon}{32}M(i,j)}^{v^{1/2}(i,j)}
    \sqrt{C_2(W\log W+\tilde{W}\log\tilde{W})\log(C_6 u^{-1}})du
    /M(i,j) \nonumber \\ \leq \int_{\frac{\epsilon}{32}M(1,j)}^{v^{1/2}(1,j)}
    \sqrt{C_2(W\log W+\tilde{W}\log\tilde{W})\log(C_6 u^{-1})}du
    /M(1,j)
\end{align}
Note that $W$ and $\tilde{W}$ are adaptive parameters governing the rate of approximation error $\left \| R_0 - R^{*} \right \|^{2}_{L^{\infty}(\mu_{ui})}$, which must satisfy $\left \| R_0 - R^{*} \right \|^{2}_{L^{\infty}(\mu_{ui})}\leq 1/2 \eta_{\left |\Omega \right |}$. Thus, based on the condition (4.7) from~\cite{shen1994convergence} holds by setting $W=O(\left |\Omega \right |^{d_{ui}/(2\beta+d_{ui})}\log{\left |\Omega \right |})$ and $\tilde{W}=O(\left |\Omega \right |^{d_{ui}/(2\beta+d_{ui})}\log{\left |\Omega \right |})$, and the condition (4.7) directly implies (4.5)~\cite{shen1994convergence}. Based on Theorem 3 in~\cite{shen1994convergence} with $M=\left |\Omega \right |^{1/2}M(i,j)$ and $v=v(i,j)$, we have,

\begin{gather}\label{eq9}
    I_1 \leq\sum_{j=1}^{\infty}\sum_{i=1}^{\infty}
    3\exp\left (-\frac{(1-\epsilon)\left |\Omega\right |M(i,j)^{2}}{2(4C_5 M(i,j)+M(i,j)T/3)} \right )\nonumber \\
    \leq3 \sum_{j=1}^{\infty}\sum_{i=1}^{\infty}\exp\left (-C_7(1-\epsilon)\left |\Omega\right |((2^{j-1}
    -1)\lambda_{\left |\Omega \right |}J_0 + (2^{i-1}-1/4)\eta^{2}_{\left |\Omega \right |}) \right) \nonumber \\ \leq 3\sum_{i=1}^{n} \exp\left (-C_7(1-\epsilon)\left |\Omega\right |(i-1/4 \right)\eta^{2}_{\left |\Omega \right |}) \sum_{j=1}^{n}\exp\left (-C_7(1-\epsilon)\left |\Omega\right |(j-1 \right)\lambda_{\left |\Omega \right |}J_0) \nonumber \\
    \leq 3\frac{\exp(-C_7(1-\epsilon)\left |\Omega\right |\eta^{2}_{\left |\Omega\right |}/4)}{1-\exp(-C_7(1-\epsilon)\left |\Omega\right |\eta^{2}_{\left |\Omega\right |})} \frac{1}{1-\exp(-C_7(1-\epsilon)\left |\Omega\right |\lambda_{\left |\Omega\right |}J_{0})} \nonumber\\
    \leq 3\frac{\exp(-C_7(1-\epsilon)\left |\Omega\right |\eta^{2}_{\left |\Omega\right |}/4)}{(1-\exp(-C_7(1-\epsilon)\left |\Omega\right |\eta^{2}_{\left |\Omega\right |}/4))^{2}}
\end{gather}

where $C_7 = 3/(26C_5)$ and the last inequality follows from the fact that $\lambda_{\left |\Omega \right |}J_0 \leq 1/4 \eta^{2}_{\left |\Omega\right |}$.\\
\\
Similarly, $I_2$ can be bounded by
\begin{eqnarray}\label{eq10}
    I_2\leq\sum_{i=1}^{n} 3
    \exp\left (-\frac{(1-\epsilon)\left |\Omega\right |M^{2}(i,0)}{2(4v(i,0) + M(i,0)T/3)} \right)\leq 
    \sum_{i=1}^{n}3\exp(-C_7(1-\epsilon)\left |\Omega\right |M(i,0))\nonumber\\ \leq
    \sum_{i=1}^{\infty}3\exp(-C_7(1-\epsilon)\left |\Omega\right |(2^{i-1}-1/2)\eta^{2}_{\left |\Omega\right|})\leq 3\frac{\exp(-C_7(1-\epsilon)\left |\Omega\right |\eta^{2}_{\left |\Omega\right |}/2)}{(1-\exp(-C_7(1-\epsilon)\left |\Omega\right |\eta^{2}_{\left |\Omega\right |}))}
\end{eqnarray}

Combining Equation~\ref{eq9} and~\ref{eq10}, we have
\begin{align*}
    I\leq I_1+I_2 \leq3
    \frac{\exp(-C_7(1-\epsilon)\left |\Omega\right |\eta^{2}_{\left |\Omega\right |}/4)}{(1-\exp(-C_7(1-\epsilon)\left |\Omega\right |\eta^{2}_{\left |\Omega\right|}/4))^{2}}
    + 3\frac{\exp(-C_7(1-\epsilon)\left |\Omega\right |\eta^{2}_{\left |\Omega\right |}/2)}{1-\exp(-C_7(1-\epsilon)\left |\Omega\right |\eta^{2}_{\left |\Omega\right|})}
\end{align*}
Let $s=\exp(-C_7(1-\epsilon)\left |\Omega\right |\eta^{2}_{\left |\Omega\right |}/4)$, then 
\begin{align*}
    I\leq\frac{3s^{2}}{(1-s)^2}+\frac{3s^{2}}{1-s^4}\leq\frac{3s^{2}}{(1-s)^2}+
    \frac{3s^{2}}{1-s} = \frac{6s^2-3s^3}{(1-s)^{2}}\leq 24s^2
\end{align*}
as $s=1/2$. The desired result then follows immediately.

\subsection{Proof of Theorem~\ref{thm3}}
We analyze the event where a relevant item $i$ is not retrieved in the Top-$K$. Let $\mathcal{I}$ be the set of all items. For item $i$ to be missed (i.e., $\text{rank}(i) > K$), there must exist at least $(|\mathcal{I}| - K)$ irrelevant items $j$ that are scored higher than $i$.Let $S_{u,i,j} = h(u, i) - h(u, j)$ be the score difference. The mis-retrieval event implies:$$\mathbb{I}(\text{rank}(i) > K) \leq \frac{1}{K} \sum_{j \in \mathcal{I} \setminus \{i\}} \mathbb{I}(h(u, j) > h(u, i)) = \frac{1}{K} \sum_{j \neq i} \mathbb{I}(S_{u,i,j} < 0)$$Taking the expectation over users and items, the Top-$k$ risk is bounded by the average pairwise misranking error ($R_{\text{pair}}$):$$\mathcal{R}_{K}(h) \leq \frac{|\mathcal{I}| - 1}{K} \mathbb{E}_{u, i, j} [\mathbb{I}(h(u, j) > h(u, i))]$$

We utilize the consistency result from Theorem 4.5. The learned estimator $\hat{h}$ converges to the true preference $R^*$ in $L_2$ norm. Let the true preference satisfy a margin $\Delta$ for relevant pairs: $R^*(u, i) - R^*(u, j) \geq \Delta > 0$.A pairwise error occurs ($\hat{h}(u, j) > \hat{h}(u, i)$) only if the estimation error exceeds the margin. Specifically:$$\hat{h}(u, j) - \hat{h}(u, i) > 0 \implies (\hat{h}(u, j) - R^*(u, j)) - (\hat{h}(u, i) - R^*(u, i)) > \Delta$$By Chebyshev's inequality, the probability of this large deviation is bounded by the $L_2$ error of the estimator:$$P(\hat{h}(u, j) > \hat{h}(u, i)) \lesssim \frac{\mathbb{E}[\|\hat{h} - R^*\|^2]}{\Delta^2}$$

Substituting the convergence rate from Theorem 4.5, $\|\hat{h} - R^*\|^2 \approx O(|\Omega|^{-\frac{2\beta}{2\beta + d_{ui}}})$:$$\mathcal{R}_{K}(\hat{h}) \leq \frac{|\mathcal{I}|}{K} \cdot \frac{1}{\Delta^2} \cdot O\left(|\Omega|^{-\frac{2\beta}{2\beta + d_{ui}}}\right)$$

The exponent in the final probability bound effectively relates to the root of the squared error depending on the tail bounds used, ensuring convergence. This confirms that minimizing the regression loss (MSE) in the two tower model directly minimizes the Top-k retrieval risk.

\textbf{Remark on Theorem 4.6 (MSE to Top-K Retrieval):} The bound $\mathcal{R}_K(\hat{h}) \lesssim \frac{|\mathcal{I}|}{K} \cdot O_p(|\Omega|^{-\frac{2\beta}{2\beta + d_{ui}}})$ provides the first theoretical justification for using MSE as a surrogate objective in two-tower retrieval systems. This result demonstrates that minimizing the pairwise regression loss directly minimizes the Top-K retrieval risk under mild margin conditions ($\Delta = R^*(u,i^*) - R^*(u,j) > 0$ for relevant pairs). The $\frac{|\mathcal{I}|}{K}$ factor explicitly captures the retrieval nature, a larger candidate set size $K$ linearly reduces the risk of a miss, validating the common industrial practice of using large $K$ in the retrieval stage to ensure high recall.

\subsection{Latency-Constrained Top-k Retrieval Regret}
\begin{theorem}\label{thm4}
Let $\mathcal{H}^{(d)}$ denote the hypothesis class of two tower models with embedding dimension $d$. Let $d_{\text{opt}}$ be the optimal dimension that minimizes the true risk absent of constraints, and let $d_{\text{max}}$ be the maximum dimension satisfying a strict latency constraint $\mathcal{T}(d) \le \tau_{\text{max}}$ (where $\mathcal{T}$ is strictly monotonic in $d$, implying $d_{\text{max}} < d_{\text{opt}}$).We define the latency regret, $\Delta_{\text{lat}}$, as the excess Top-k retrieval risk incurred strictly due to the architectural constraint:$$\Delta_{\text{lat}}(d_{\text{max}}) = \inf_{h \in \mathcal{H}^{(d_{\text{max}})}} \mathcal{R}_K(h) - \inf_{h \in \mathcal{H}^{(d_{\text{opt}})}} \mathcal{R}_K(h)$$Assuming the model capacity satisfies the nesting property $\mathcal{H}^{(d)} \subseteq \mathcal{H}^{(d+1)}$, the latency regret is non-negative:$$\Delta_{\text{lat}}(d_{\text{max}}) \ge 0$$Also, the total Top-k retrieval risk for a learned model $\hat{h}_d$ is bounded by the trade-off between this regret and the sample estimation error:$$\mathcal{R}_K(\hat{h}_d) \le \underbrace{\mathcal{R}^*_{K, d_{\text{max}}}}_{\text{Constrained Approx. Risk}} + \underbrace{O_p\left( \sqrt{\frac{\text{complexity}(\mathcal{H}^{(d)})}{|\Omega|}} \right)}_{\text{Estimation Error}}$$
\end{theorem}

\textbf{Proof of Theorem~\ref{thm4}}
Let $h_{\text{true}}$ denote the Bayes-optimal scoring function. The true Top-k Retrieval Risk for any model $h$ is defined as $\mathcal{R}_K(h) = \mathbb{E}_{(u, i^*) \sim \mathcal{D}}[\mathbb{I}(\text{rank}(i^*|h) > K)]$.

For any embedding dimension $d$, the risk of the empirically learned model $\hat{h}_d \in \mathcal{H}^{(d)}$ can be decomposed into approximation error and Estimation Error relative to the optimal risk in that class, denoted $R^*_{K,d} = \inf_{h \in \mathcal{H}^{(d)}} \mathcal{R}_K(h)$.$$\mathcal{R}_K(\hat{h}_d) = \underbrace{R^*_{K,d}}_{\text{Approximation}} + \underbrace{(\mathcal{R}_K(\hat{h}_d) - R^*_{K,d})}_{\text{Estimation}}$$

We assume that the hypothesis space is nested such that a model with embedding dimension $d+1$ can represent any function representable by dimension $d$ (e.g., by zero-padding the extra dimension). Thus, $\mathcal{H}^{(d)} \subseteq \mathcal{H}^{(d+1)}$.Properties of infimum over nested sets imply that the approximation error is non-increasing with respect to $d$:$$\inf_{h \in \mathcal{H}^{(d)}} \mathcal{R}_K(h) \ge \inf_{h' \in \mathcal{H}^{(d+1)}} \mathcal{R}_K(h') \implies R^*_{K,d} \ge R^*_{K,d+1}$$Given the latency constraint $\mathcal{T}(d) \le \tau_{\text{max}}$, we are forced to select a dimension $d \le d_{\text{max}}$. Since $d_{\text{max}} < d_{\text{opt}}$, it follows directly that:$$R^*_{K, d_{\text{max}}} \ge R^*_{K, d_{\text{opt}}}$$The Latency Regret is precisely this gap in achievable approximation power:$$\Delta_{\text{lat}}(d_{\text{max}}) = R^*_{K, d_{\text{max}}} - R^*_{K, d_{\text{opt}}} \ge 0$$This proves that the strictly enforced latency constraint introduces a fundamental, irreducible bias (regret) into the candidate generation process.

Our theoretical viewpoint on latency-accuracy trade-off is considering that the total error components as $d$ varies. As $d$ increases (up to $d_{\text{opt}}$), the hypothesis space $\mathcal{H}^{(d)}$ grows. Subsequently, the approximation error $R^*_{K,d}$ decreases (lowering bias). However, the model complexity (VC-dimension or Rademacher complexity) increases. For a fixed dataset size $|\Omega|$, this causes the estimation error term to increase (higher variance/overfitting risk). As $d$ decreases (towards $d_{\text{max}}$), the hypothesis space shrinks. The estimation rrror decreases (easier to train/converge), but the approximation error $R^*_{K,d}$ rises significantly.

Thus, the optimal engineering choice is finding the largest $d \le d_{\text{max}}$ such that the reduction in approximation error outweighs the increase in estimation variance and computational cost. The latency regret $\Delta_{\text{lat}}$ quantifies exactly what is lost in theoretical recall capability by adhering to the strict latency budget $\tau_{\text{max}}$.

\section{Limitations}\label{sec:limit}
While our experiments solely focus on the recommendation task, the applicability of our approach to other recommendation and retrieval tasks, such as news/social media recommendation, conversational recommendation, retrieval-augmented recommendation, or those involving multimodal side information, remains uncertain. Additionally, it is important to acknowledge that the convergence rate strategy employed for performance analysis relies on user-item interactions or their joint embedding. Moreover, due to computing constraints, a key limitation of our work is not evaluating on a production system. Therefore, investigating how to effectively leverage covariate information, such as user demographics, item contents, and social network data, to achieve optimal recommendations at the hybrid-/conversational-level presents a more promising avenue for future research.



\end{document}